\documentclass[final,3p,times,twocolumn]{elsarticle}

\usepackage{amssymb}
\usepackage{amsmath}

\newcommand{\siesta}{\textsc{SIESTA~}}

\newcommand{\dftbp}{\textsc{DFTB+~}}
\newcommand{\bsigma}{\boldsymbol{\sigma}}

\usepackage{appendix}
\usepackage[final, authormarkup=none]{changes}
\usepackage{graphicx}
\usepackage{dcolumn}
\usepackage{bm}
\usepackage{hyperref}
\biboptions{sort&compress}

\definechangesauthor[name={Per cusse}, color=green]{per}
\definechangesauthor[name={Alex}, color=orange]{AC}
\definechangesauthor[name={Aleksander}, color=red]{ABL}
\definechangesauthor[name={Aleks}, color=green]{ABL2}
\definechangesauthor[name={Aleks}, color=purple]{ABL3}
\definechangesauthor[name={Aleks}, color=green]{ABL4}

\definechangesauthor[name={MB}, color=purple]{MB}

\journal{Computer Physics Communications}

\begin{document}

\begin{frontmatter}

\title{Zandpack:\\A General Tool for Time-dependent Transport
Simulation of Nanoelectronics}

\author[dtu]{Aleksander Bach Lorentzen}
\author[jena]{Alexander Croy}
\author[dtu]{Antti-Pekka Jauho}
\author[dtu]{Mads Brandbyge}
\affiliation[jena]{
 Institute of Physical Chemistry, Friedrich-Schiller-Universität Jena, 07737 Jena, Germany
}
\affiliation[dtu]{
             organization={Physics Department, Technical University of Denmark},
             city={Kongens Lyngby},
             postcode={2800},
             country={Denmark}
}
\ead{mabr@dtu.dk}

\date{\today}

\begin{abstract}
The auxiliary mode approach to time-dependent open quantum system calculations is implemented and refined to yield a feasible computational approach to simulate nanostructures far from equilibrium. It is done by a careful diagonalization of the electrode level-width function, and provides an efficient approach which can simulate large, open systems at the level of time-dependent density functional theory. The approach, as given in this work, is implemented in the new open-source code Zandpack. The framework is applied to three systems perturbed by the same THz electromagnetic field pulse-form: 1) A Hubbard model for hydrogen on graphene is used to calculate spin-currents, mutual information, spin-transitions, and a pump-probe setup. 2) An armchair graphene nanoribbon (AGNR) probed by a metal tip showing electrons excited from the valence band of the AGNR into the tip via electron-electron interactions. 3) A gold break-junction is modeled with various gap distances, and displays behavior that is more different from the adiabatic case as the gap widens. In the examples 2 and 3, we develop and use a general linearization scheme for time-dependent open system calculations, which utilizes the \dftbp or \siesta codes.

\noindent \textbf{PROGRAM SUMMARY}
\noindent \\
\noindent \\
{\em Program Title: }Zandpack                                          \\
{\em CPC Library link to program files:} (to be added by Technical Editor) \\
{\em Developer's repository link:} https://github.com/AleksBL \\
{\em Licensing provisions:} MPL-2.0  \\
{\em Programming language: } Python                                  \\
{\em Nature of problem:}\\
Simulating the evolution of electrons in a device region connected to electrodes that are experiencing a time-dependent and strong bias, while at the same time describing the evolving device electrons accurately at the mean-field level. The electronic structure of the electrodes must also be described accurately in terms of the energy-dependence of its level-width function. This on a practical level requires fitting a set of known functions to a sum of Lorentzians. This fit then fixes the parameters of a coupled system of ODEs, in which the electronic density and Hamiltonian appears. Lastly, this system of ODEs has to be solved numerically.
\\
{\em Solution method:}\\
A user-friendly routine for fitting the electrode level-width functions is implemented. It can either take input from TBtrans or custom user input and convert it to a sum of Lorentzians. We employ the auxiliary mode expansion (AME) method following Popescu and Croy [New J. Phys. 18, 093944 (2016)] with a modified version of the diagonalization technique, combined with an effective account for the electrode level-width functions. The code can obtain the initial steady state, and propagate this initial steady state after application of a user-defined voltage bias-pulse applied to the electrodes using an explicit Runge-Kutta solver. Throughout this propagation, a user-defined density-dependence is needed, e.g. by interfacing to an external LCAO-DFT code. Such an interface is available for SIESTA and DFTB+, but can also be written by the user.
\\
{\em Additional comments including restrictions and unusual features:}\\
The AME method does not have any restrictions on how fast oscillations can be, meaning it is valid for slow (e.g. THz fields) as well as fast (e.g. optical fields) perturbations. Simulations with normal-superconducting-normal type setups are also possible.
In the current implementation, phonons cannot be included in the calculation, but the method does in theory allow for such [Y. Zhang, C. Y. Yam, G. Chen, J. Chem. Phys. 138 (16) (2013)].
\end{abstract}

\begin{highlights}
 \item A code implementing the auxiliary mode approach for time-dependent quantum transport in open systems is presented. The code is demonstrated using three examples of nano-scale devices applying different levels of complexity in the mean-field description of the electronic structure.
\end{highlights}


\end{frontmatter}

\newcommand{\Packagename}{zand} 
\newcommand{\FitPackageName}{Zandpack} 
\makeatletter
\DeclareRobustCommand\widecheck[1]{{\mathpalette\@widecheck{#1}}}
\def\@widecheck#1#2{%
    \setbox\z@\hbox{\m@th$#1#2$}%
    \setbox\tw@\hbox{\m@th$#1%
       \widehat{%
          \vrule\@width\z@\@height\ht\z@
          \vrule\@height\z@\@width\wd\z@}$}%
    \dp\tw@-\ht\z@
    \@tempdima\ht\z@ \advance\@tempdima2\ht\tw@ \divide\@tempdima\thr@@
    \setbox\tw@\hbox{%
       \raise\@tempdima\hbox{\scalebox{1}[-1]{\lower\@tempdima\box
\tw@}}}%
    {\ooalign{\box\tw@ \cr \box\z@}}}
\makeatother

\section{\label{sec:intro} Introduction}
During the last decade scanning near-field optical microscopy (SNOM) and its THz equivalent have been developed and enhanced to utilize a scanning tunneling microscope (STM) by several experimental groups \cite{Cocker2013, yoshida2015terahertz, li2024real, wang2022atomic, maier2025attosecond}. SNOM is firstly a measurement where the apex of an atomic force microscope (AFM) tip is brought into proximity of a surface to be probed, after which a THz/IR pulse is shone upon the tip, giving a significant enhancement of the electric field just below the tip. This allows probing the material just below the tip with a resolution on the nanometer scale \cite{Chen2019}. This idea of using a strong THz field and a tip to probe the material in question has since been refined to give atomic resolution by using field-driven tunneling from the material to a STM-tip (THz-STM) \cite{garg2022real, Ammerman2022,ammerman2021lightwave,Cocker2021, Yoshida2021, nguyen2020modeling, Luo2020, Jelic2017, Cocker2013, yoshida2015terahertz, li2024real}. Both SNOM and THz-STM open up the realm of non-equilibrium phenomena on the scale where quantum effects are significant. The THz pulse can furthermore be combined with higher frequency pulsed light to make a pump-probe setup, where a weaker near-optical or optical pulse perturbs and probes the system at a variable time during the THz pulse \cite{Cocker2013, Ammerman2022, liang2023ultrafast, Yoshida2021}. This type of experiment in particular combines the relatively short timescale of the optical pulse with the longer THz pulse, which makes for a theoretically challenging task as the simulation method used must handle both the optical transitions in the junction, together with the slowly varying field from the THz pulse.

Simulations for the time-independent case have been standard in the nano-electronics community for many years now. Different techniques for efficient evaluation of the non-equilibrium steady state density matrix have been developed \cite{papior2017improvements, stokbro2003transiesta, brandbyge2002density,garcia2020siesta, kohn1965self}. Non-equilibrium Green's function (NEGF) theory for dealing with time-dependent electronic transport in non-equilibrium conditions is a well-established field, albeit computationally heavy. It can model time-dependent phenomena in open single particle systems, systems with many-body interactions and ``normal-superconducting-normal''-type systems \cite{jauho1994time, haug2008quantum, Zheng2007, Jin2008, zheng2010time, ridley2022many,Popescu2016, xie2012time, hu2011pade, Croy2009, Wang2019, Beltako2019, Wang2021, Hu2021, tuovinen2016time, tuovinen2016article, bergmann2021green, seshadri2021entropy, Hirsbrunner2019, honeychurch2019timescale, wang2022theoretical,wang2015time}.

Software for simulation of nanoelectronics is already available in various packages, such as NESSi, Cheers, TKwant and HEOM-QUICK \cite{schuler2020nessi, kloss2021tkwant, ye2016heom, pavlyukh2023cheers}, which implement the NEGF equations using a range of different approximations. These do have various difficulties associated with them: NESSi is a general many-body code which is computationally expensive. Tkwant is a wavefunction-based code which means the density matrix needs to be computed at each time instance and also introduces the embedding self-energies which will follow in Sec. 2 in a different way\cite{kloss2021tkwant}. The driven Liouville-von Neumann (DLvN) approach is also a method applicable to the same problem, but it is phenomenological in nature\cite{hod2023driven, zelovich2017parameter} and not NEGF based. Common to NEGF codes is that the single particle density matrix is available from the lesser Green's function ($G^<$). The electronic density can be obtained directly from the density matrix, meaning NEGF based methods are compatible with density functional theory (DFT). The body of available codes implementing DFT in various ways is large (see Ref.\ \cite{wikidftcodes}), and namely, the codes using localized basis functions are of particular use in quantum transport. For this paper, the DFTB+\cite{aradi2007dftb+, hourahine2020dftb+} and SIESTA\cite{garcia2020siesta, soler2002siesta} codes have been used to describe the electronic structure. Zandpack itself is agnostic to the particular DFT code used, as long as a Python wrapper function can be made. 

A single-particle density-matrix-based NEGF method, which is computationally efficient, has however been developed in Ref.\ \cite{Popescu2016}. It can be combined with DFT and is applicable to larger systems. The strength of the method at hand is that it allows for an exact analytical treatment - within the level-width function fitting scheme and independent particle picture - of any bias-pulse of and electrode with an arbitrary electronic structure. This allows for \textit{ab-initio} time-dependent transport calculations, as the electrode level-width functions, $\mathbf{\Gamma}_\alpha$, can be fitted to the needed precision. It is furthermore extensible to Kohn-Sham TDDFT because the electron density is readily available from the density matrix. Dynamical effects can therefore also be included through a retarded response function\cite{li2020real}.
In this paper we describe a computationally efficient implementation of the method and clarify subtle, but important requirements for obtaining physically sensible results using the method. The described functionality is available in the new python-based software package Zandpack together with Jupyter notebooks demonstrating its use \cite{zandpack_web}.

Zandpack contains the necessary tools to fit the electrode level-width functions to a sum of Lorentzians, together with tools for finding the steady state and carrying out the time-propagation. The provided software implementation can only deal with electronic correlations in the mean-field picture, but there are extensions of the methodology to perturbatively include electron-phonon interactions \cite{chen2018time, zhang2013dissipative}. It does not deal with strong correlation effects, e.g. as is found in a Mott insulating phase.

\section{\label{sec:theory}Theory}
\subsection{The Transport Setup}
\begin{figure}[h]
    \centering\includegraphics[width=5.0cm]{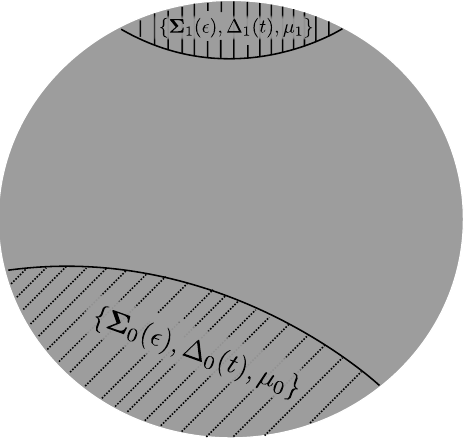}
    \caption{Device region of a transport setup. Regions 0 and 1 denoted on the figure are where the device region (gray) couple to the surrounding environment.}
    \label{fig:TransportSetup}
\end{figure}

In transport theory, the basic setup is the electrode-device setup (usually a left-device-right setup), in which the electrodes are described by a self-energy, $\mathbf{\Sigma}$, defined inside the device region, as illustrated in Fig.\ \ref{fig:TransportSetup}.
 In the framework of  NEGF, the self-energy $\mathbf{\Sigma}$ introduces the effects of the infinite environment through an energy-dependent matrix-valued function $\mathbf{\Sigma}(\epsilon)$, in a numerically exact way\cite{haug2008quantum, ridley2022many}. The description of electrodes is completed by a chemical potential $\mu_\alpha$ and $\Delta_\alpha(t)$, which is a time-dependent bias that describes a rigid, time-dependent shift of the electronic structure of electrode $\alpha$. Additionally, thermal energies, $k T_\alpha$, in the electrodes are needed in addition to the indicated quantities in Fig.\ \ref{fig:TransportSetup} to obtain the description at finite temperature.

\subsection{Non-equilibrium Green's Functions}
The primary quantity of interest in electronic structure theory and time-dependent calculations is the electronic density matrix, which is a derived quantity from the lesser Green function of the system. The equation of motion of the density matrix for the full system is given by the usual Liouville-von Neumann equation of motion \cite{Zheng2007,rammer1991quantum, haug2008quantum}
\begin{align}
    \label{eq:QLoui}
    i\hbar\frac{\mathrm{d}}{\mathrm{d}t}{\bsigma}_F(t)= [\mathbf{H}_F(t),  {\bsigma}_F(t)]\,, 
\end{align}
where $\mathbf{H}_F$ is the Hamiltonian of the full system, and $\bsigma_F(t)= -i\hbar \mathbf{G}_F^<(t,t)$ is the density matrix of the full system: 
\begin{align}
  \mathbf{H}_{F} =
  \left[ {\begin{array}{ccc}
    \mathbf{H}_{L} & \mathbf{H}_{L, D} & 0\\
    \mathbf{H}_{D,L} & \mathbf{H}_{D}&\mathbf{H}_{D,R} \\
    0 & \mathbf{H}_{R,D} & \mathbf{H}_{R}
  \end{array} } \right],
\end{align}
and
\begin{align}
  \bsigma_{F} =
  \left[ {\begin{array}{ccc}
    \bsigma_{L} & \bsigma_{L, D} & \bsigma_{LR}\\
    \bsigma_{D,L} & \bsigma_{D}&\bsigma_{D,R} \\
    \bsigma_{R,L} & \bsigma_{R,D} & \bsigma_{R}
  \end{array} } \right]\,,
\end{align}
in the case of the left-device-right setup and can be generalized to a $N$-electrode setup. The Liouville-von Neumann equation is intractable for a system with electrodes that have an effectively infinite extent. However, by considering only the elements inside a restricted region, the device region, it is possible to obtain the evolution of the density matrix of this sub-system as \cite{Zheng2007, Croy2009, Popescu2016, ridley2022many}
\begin{align}
    \label{eq:sigEOM}
    i\hbar\frac{\mathrm{d}}{\mathrm{d}t}\bsigma_{D}(t)= [\mathbf{H}_{D}(t),  \bsigma_{D}(t)] + i\sum_{\alpha} \left[\mathbf{\Pi}_\alpha(t) + \mathbf{\Pi}_\alpha^\dagger(t)\right]\,,
\end{align}
where the current matrices, $\mathbf{\Pi}_\alpha$, account for the effects of the $n_\alpha$ uncoupled electrodes in the device region while keeping the problem tractable. The current matrices are given as,
\begin{align}
    \label{eq:PiForm}
    \mathbf{\Pi}_{\alpha}(t) &= \int_{t_0}^t\mathrm{d}t'\left[\mathbf{G}_D^>(t, t')\mathbf{\Sigma}^<_{\alpha}(t' , t) - \mathbf{G}_D^<(t, t')\mathbf{\Sigma}^>_{\alpha}(t' , t)\right],
\end{align}
and they yield the current flowing from the electrode $\alpha$ as $J_\alpha(t) = \frac{2\mathrm{e}}{\hbar}\mathrm{Re}\mathrm{Tr}\left[\mathbf{\Pi}_\alpha(t)\right]$. From here on the "$D$"-subscript will be dropped. 
The electrode self-energy components $\mathbf{\Sigma}_{ \alpha}^{</>}$ in Eq.\ \eqref{eq:PiForm} are generalized scattering rates out of / into the device, and are given as 
\cite{Croy2009, Popescu2016,jauho1994time, haug2008quantum}, 
\begin{align}
    \label{eq:L_SE}
    \mathbf{\Sigma}^<_{\alpha}(t_2,t) = &\phantom{-}\frac{i}{2\pi\hbar}\mathrm{e}^{\frac{i}{\hbar}\int_{t_2}^t \mathrm{d}t_1 \Delta_\alpha(t_1)}\\
    &\times\int_{-\infty}^\infty\mathrm{d}\epsilon f_\alpha(\epsilon)\mathbf{\Gamma}_{ \alpha}(\epsilon)\mathrm{e}^{-\frac{i}{\hbar}\epsilon(t_2 - t)}, \nonumber\\
    \label{eq:G_SE}
     \mathbf{\Sigma}^>_{ \alpha}(t_2,t) = &-\frac{i}{2\pi\hbar}\mathrm{e}^{\frac{i}{\hbar}\int_{t_2}^t \mathrm{d}t_1 \Delta_\alpha(t_1)}\\
     &\times\int_{-\infty}^\infty\mathrm{d}\epsilon (1-f_\alpha(\epsilon))\mathbf{\Gamma}_{ \alpha}(\epsilon)\mathrm{e}^{-\frac{i}{\hbar}\epsilon(t_2 - t)},\nonumber
\end{align}
where $f_\alpha(\epsilon)=1/(\exp((\epsilon-\mu_\alpha)/kT_\alpha)+1)$ is the Fermi-Dirac distribution of the electrode, $\Delta_\alpha$ the time-dependent electrode bias, and $\mathbf{\Gamma}_\alpha$ is the electrode level-width function.
Note that here it is assumed that in the electrode regions (dashed in Fig. \ref{fig:TransportSetup}) the state energies {\em and} fillings rigidly follow the electrode bias shift $\Delta_\alpha(t)$.
The retarded/advanced components of the self-energy can furthermore be obtained by the relation \cite{haug2008quantum},
\begin{align}
    \mathbf{\Sigma}_\alpha^{r/a}(t_2, t) = \pm \Theta(\pm t_2 \mp t)\left[\mathbf{\Sigma}_\alpha^>(t_2, t) - \mathbf{\Sigma}_\alpha^<(t_2, t)\right],
\end{align}
{where $\Theta$ is the Heaviside step function. The retarded self energy, $\mathbf{\Sigma}_\alpha^r$, contains spectral information about electrode $\alpha$.
\subsection{Pole Expansion}
The electrode level-width functions, which are obtained from the bare electrode Green's function $\mathbf{g}_\alpha^r$, 
\begin{align}
    \mathbf{\Gamma}_\alpha= \mathbf{H}_{D\alpha}(\mathbf{g}_\alpha^r - \mathbf{g}_\alpha^a)\mathbf{H}_{\alpha D} = i\left[\mathbf{\Sigma}^r_\alpha - \mathbf{\Sigma}^a_\alpha\right]
\end{align}
are assumed to be given by a weighted sum of Lorentzian functions\cite{Croy2009,Popescu2016}:
\begin{align}
\label{eq:GammaForm}
    \mathbf{\Gamma}_{\alpha}(\epsilon) &= \sum_l^{N_l} L_{\alpha l }(\epsilon) \mathbf{W}_{\alpha l}\,,
\end{align}
where $\mathbf{W}_{\alpha l}$ is a matrix of coefficients and 
\begin{align}
\label{eq:Lorentzian}
L_{\nu}(\epsilon) = \frac{\gamma_\nu ^2 }{(\epsilon - \epsilon_\nu)^2 + \gamma_\nu^2} = \frac{i\gamma_\nu}{2}[\frac{1}{\epsilon-z_\nu^-} - \frac{1}{\epsilon - z_\nu^+}]\,,
\end{align}
is a Lorentzian function with $z_\nu^+=(z_\nu^-)^* =  \epsilon_\nu + i\gamma_\nu$. The coefficient matrix $\mathbf{W}_{\alpha l}$ furthermore has the eigendecomposition
\begin{align}
    \label{eq:LambdaEig}
    \mathbf{W}_{\alpha l} &= \sum_c \lambda_{\alpha l c}\,\vec{\xi}_{\alpha l c}\otimes\vec{\xi}^{\phantom{.}\dagger}_{\alpha l c}\,,
\end{align}
where the summation index $c$ runs over the number of non-zero eigenvalues of $\mathbf{W}_{\alpha l}$. This form is just the regular $\mathbf{M} = \mathbf{U}\mathbf{D}\mathbf{U}^\dagger$ eigendecomposition written as an explicit sum. The $\mathbf{W}_{\alpha l}$ matrices in the above sum can be fitted with the python package \FitPackageName \phantom{a}to \textit{ab-initio} level-width functions calculated using, e.g., the TBtrans code\cite{papior2017improvements}. At this point we have to introduce the {\em energy window} around the initial, equilibrium Fermi energy. All essential states involved in the time-dependent dynamics, and the corresponding poles, should be located within this energy window on the real energy axis. Clearly, the size of the window must extend beyond the peak electrode potentials ($\Delta_\alpha(t) + \mu_\alpha$), but the concrete window size is a matter of case-by-case convergence (cf. App.~\ref{app:convergence}).

The fit of $\mathbf{\Gamma}_\alpha$'s to the numerically exact broadening matrices $\tilde{\mathbf{\Gamma}}_\alpha = i(\tilde{\mathbf{\Sigma}}_\alpha-\tilde{\mathbf{\Sigma}}_\alpha^\dagger)$ is the first part of the numerical procedure of going from the standard, time-independent  transport setup, to the auxiliary mode time-propagation. Importantly, $\tilde{\mathbf{\Gamma}}_\alpha(\epsilon)$ is positive semi-definite for all $\epsilon \in \mathbb{R}$ and the fitted $\mathbf{\Gamma}_{\alpha}(\epsilon)$ has to be so too \cite{wang2015time}. For this purpose, the Zandpack code is used to determine the coefficients $\mathbf{W}_{\alpha l}$ from Eq.\ \eqref{eq:GammaForm}. The fitting procedure is elaborated on in App.~ \ref{App:fitting}.

It should also be noted that the form in Eq.\ \eqref{eq:GammaForm} also determines the retarded self-energy through the relation
\begin{align}
\label{eq:HilbertSE}
    \mathbf{\Sigma}_{\alpha}^r (\epsilon) = \frac{1}{2}\mathcal{H}[\mathbf{\Gamma}_{\alpha}](\epsilon)  -\frac{i}{2}\mathbf{\Gamma}_{\alpha}(\epsilon)\,,
\end{align} 
where $\mathcal{H}$ denotes the Hilbert transform\cite{papoulis1963fourier}. The replacement of  $\mathbf{\Gamma}_\alpha$ by an approximation with support limited to the energy window, will show up in the real (Hilbert) part of $ \mathbf{\Sigma}_{\alpha}^r$. This is illustrated in Fig.~\ref{fig:HilbertDemon} for a simple example. Here we show how the Hilbert transform inside the energy window (white parts) can be impacted by the neglected real part of $\Sigma^r$ outside the energy window (gray parts).

\begin{figure}[h]
    \centering
    \includegraphics[width=7.0cm]{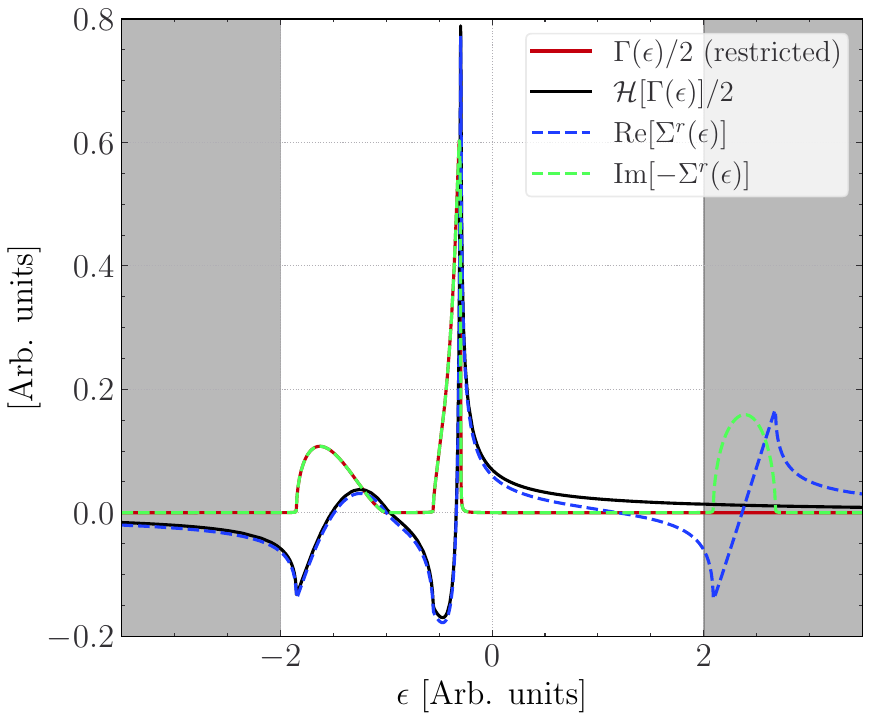}
    \caption{The impact on the Hilbert transform (see Eq. \eqref{eq:HilbertSE}) of a function $\Gamma = i\left[\Sigma^r - \Sigma^{r\dagger}\right]$ that has been set to zero for $|\epsilon|>2$. Here it is compared to its unrestricted counterpart $-\mathrm{Im}\Sigma^r$. Grey indicates the energy window where $\Gamma$ is assumed to be zero, and $\Sigma^r$ is the full self-energy.}
    \label{fig:HilbertDemon}
\end{figure}

Furthermore, doing a sum-over-poles expansion for the scaled Fermi-function of the form \cite{croy2009propagation,hu2010communication,Popescu2016},
\begin{align}
\label{eq:FermiFunc}
    F(s) =\frac{1}{1 + \mathrm{e}^s} = \frac{1}{2} -  \sum_p\frac{R_p}{s - z_p^+} + \frac{R_p}{s - z_p^-}
\end{align}
gives poles for the electrode Fermi function $f_\alpha(\epsilon) = F(\frac{\epsilon - \mu_\alpha}{kT_\alpha})$ at $\epsilon = \mu_\alpha + kT_\alpha z^{\pm}_p $.

Having the fit, the energy-integrals in Eqs.\ \eqref{eq:L_SE} and \eqref{eq:G_SE} are done using the residue-theorem and Jordan's lemma \cite{Popescu2016}. This requires a sum over the residues of the integrands in the self-energies. If the Fermi-function $f$ is also expanded in poles (as in Refs.\ \cite{hu2010communication, croy2009propagation}), then all poles of the integrand are known, namely, the union between the poles of the Lorentzian functions and the poles of the Fermi functions, $f_\alpha$. All poles are symmetrically distributed around the real axis. These poles are denoted by $\chi^{\pm}_{\alpha x}$ where $x$ is the combined index of the Fermi-poles and Lorentzian-poles, with the $\pm$ sign indicating the sign of the imaginary part of the pole. Furthermore, the quantity $\chi^{\pm}_{\alpha x} + \Delta_\alpha(t)$ will often appear, so we will introduce the following shorthand definition,
\begin{align}
    \chi^{\pm}_{\alpha x}(t) = \chi^{\pm}_{ \alpha  x} + \Delta_\alpha(t)\,.
\end{align}

The vectors $\vec{\xi}_{\alpha x c}$ will now appear with the $x$ index, extending the eigendecomposition in Eq.\ \eqref{eq:LambdaEig} to also include the eigencompositions of $\mathbf{\Gamma}_\alpha(z_p)$ in the Fermi poles. This is because the residual of the integrand must be computed in a Fermi-pole using Eq.\ \eqref{eq:GammaForm}. Additionally, $\mathbf{\Gamma}_\alpha(z_p)$ is split in the following way: 
\begin{align}
\label{eq:GammaDecomp}
    \mathbf{\Gamma}_{ \alpha}(z_p)& =\frac{\mathbf{\Gamma}_{ \alpha}(z_p) + \mathbf{\Gamma}^\dagger_{ \alpha}(z_p)}{2} + \frac{\mathbf{\Gamma}_{ \alpha}(z_p) - \mathbf{\Gamma}^\dagger_{ \alpha}(z_p)}{2} \\
    &=  \widehat{\mathbf{\Gamma}}_{ \alpha}(z_p) + \widecheck{\mathbf{\Gamma}}_{ \alpha}(z_p)\,.
\end{align}
These two matrices, contrary to their sum, are individually diagonalizable by a unitary transformation, which is a necessary requirement to obtain the set of ordinary differential equations (ODE) given below. Importantly, we note that without a unitary decomposition, we may obtain spurious currents flowing under equilibrium conditions, which is clearly non-physical. Following Ref.\ \cite{Popescu2016} now gives the self-energies as a sum of outer products:
\begin{align}
    \label{eq:SE_EigDecomp}
    \mathbf{\Sigma}_{\alpha}^{</>}(t_2, t) = \frac{1}{\hbar}\sum_{c,x}\Lambda^{</>,\pm}_{\alpha x c}\vec{\xi}_{\alpha x c}\otimes\vec{\xi}^{\phantom{.}\dagger}_{\alpha x c}\mathrm{e}^{-\frac{i}{\hbar}\int_{t_2}^t \mathrm{d}t'\chi_{\alpha x}^{\pm}(t')}\,,
\end{align}
where the $\pm$ depends on the time-ordering of the phase-factor, the index $x$ runs over $N_L$ indices corresponding to the Lorentzian poles of the $\mathbf{\Gamma}_\alpha(\epsilon)$ expansion, and $2N_F$ indices are coming from the two matrices of each symmetric-antisymmetric decomposition of the $N_F$ different $\mathbf{\Gamma}_\alpha(z_p)$. For each type of pole, the residues inside the $</>$ self-energies evaluates to,
\refstepcounter{equation}
\begin{align}
    \Lambda_{\alpha l c}^{<, \pm} &= \phantom{-}\frac{i}{2}f_\alpha(z_{\alpha l }^{\pm})\lambda_{\alpha lc}\gamma_{\alpha l}\,,  \nonumber 
    \\
    \Lambda_{\alpha \widehat{p} c}^{<, \pm} &= \pm R_{\alpha p}kT_\alpha \widehat{\lambda}_{\alpha p c}\,, \nonumber 
    \\
    \Lambda_{\alpha \widecheck{p} c}^{<, \pm} &= \pm R_{\alpha p}kT_\alpha \widecheck{\lambda}_{\alpha p c}\,,\tag{\theequation a-f}
    \\
    \Lambda^{>,\pm}_{\alpha l c} &= -\frac{i}{2}(1-f_\alpha(z_{\alpha l }^{\pm}))\lambda_{\alpha l c}\gamma_{\alpha l}\,, \nonumber 
    \\
    \Lambda_{\alpha \widehat{p} c}^{>, \pm} &= \pm R_{\alpha p} kT_\alpha \widehat{\lambda}_{\alpha p c}\,, \nonumber
    \\ 
    \Lambda_{\alpha \widecheck{p} c}^{>, \pm} &= \pm R_{\alpha p} kT_\alpha \widecheck{\lambda}_{\alpha p c}\,.\nonumber
\end{align}
Here $\widehat{\lambda}_{ \alpha p c} \phantom{n}( \widecheck{\lambda}_{ \alpha p c})$ are the accompanying eigenvalues of $\widehat{\mathbf{\Gamma}}_\alpha(z_p)  \phantom{n}( \widecheck{\mathbf{\Gamma}}_\alpha(z_p))$. Thus, the general index, $x$, now take values corresponding to the Lorentzian expansion, $l$, or a Fermi-pole requiring two indices, $\widehat{p}$ and $\widecheck{p}$, while index $c$ runs over the corresponding eigendecomposition for each of these matrices.
The $\pm$ components of the $\Lambda$-quantities are defined relative to the electrode self-energy times as $t> t_2$ ($t< t_2$) corresponding to $+$ ($-$).

It is now possible to define the auxiliary mode wave-vector $\vec{\Psi}_{\alpha x c}(t)$, that gives rise to a decomposition for $\mathbf{\Pi}_{\alpha}$ as a sum of outer products of the form \cite{Popescu2016},
\begin{align}
\label{eq:PIfromPSI}
    \mathbf{\Pi}_{\alpha}(t) = \frac{1}{\hbar} \sum_{x c}\vec{\Psi}_{\alpha x c}(t) \otimes\vec{\xi}^{\phantom{.}\dagger}_{\alpha x c}
\end{align}
by identifying
\begin{align}
    \label{eq:PsiDef}
    \vec{\Psi}_{ \alpha x c}(t) = \int_{t_0}^t\mathrm{d}t'&[\textbf{G}^>(t,t')\Lambda_{\alpha x c}^{<,+} - \textbf{G}^<(t,t')\Lambda_{\alpha x c}^{>,+}]\nonumber\\ &\times\mathrm{e}^{\frac{i}{\hbar} \int_{t'}^t\mathrm{d}t''\chi_{\alpha x c}(t'')}\vec{\xi}_{ \alpha x c}
\end{align}
from Eq.\ \eqref{eq:PiForm} and inserting Eq.\ \eqref{eq:SE_EigDecomp}.
The present method propagates the system described by the Hamiltonian, self-energies in terms of Lorentzian and Fermi functions, and the time-dependent bias in an exact way in the one-particle picture, with direct access to the density matrix. The method achieves this by introducing an auxiliary wave vector, $\vec{\Psi}_{ \alpha x c}(t)$, whose equation of motion is given in Ref.\ \cite{Popescu2016}, page 5:
\begin{align}
    i\hbar\frac{\mathrm{d}\phantom{l} }{\mathrm{d}t}\vec{\Psi}_{\alpha x c} &=[\mathbf{H}(t) -  \mathbf{1}\,\chi^+_{\alpha x }(t) ]\vec{\Psi}_{\alpha x c}  + \hbar\Lambda^{<+}_{\alpha x c}\vec{\xi}_{\alpha x c}  \label{eq:psiEOM} 
    \\
    &+\hbar(\Lambda^{>,+}_{\alpha x c} - \Lambda^{<,+}_{\alpha x c})\bsigma\vec{\xi}_{\alpha x c} \nonumber
    \\
    & + \frac{1}{\hbar}\sum_{\alpha'x'c'}\Omega_{\alpha x c \alpha' x' c' }\vec{\xi}_{\alpha'x'c'}\,,\nonumber
    \\
    i\hbar\frac{\mathrm{d}\phantom{l}}{\mathrm{d}t}\Omega_{\alpha x c \alpha' x' c' } &= (\chi^-_{\alpha' x'  }(t) - \chi^+_{\alpha x }(t))\Omega_{\alpha x c \alpha' x' c' }  \label{eq:omgEOM}
    \\
    &\phantom{=}+i\hbar(\Lambda^{>,-}_{\alpha' x'{} c'} - \Lambda^{<,-}_{\alpha' x' c'})\vec{\xi}^{\phantom{.}\dagger}_{\alpha' x' c'}\vec{\Psi}_{\alpha x c} \nonumber
    \\
    &\phantom{=}+i\hbar(\Lambda^{>,+}_{\alpha x c}\phantom{n.} - \phantom{n}\Lambda^{<,+}_{\alpha x c})\vec{\Psi}^\dagger_{\alpha' x' c'}\vec{\xi}_{\alpha x c}\,.\nonumber
\end{align}
The quantity $\Omega_{\alpha x c \alpha' x' c' } $ appears when $\frac{\mathrm{d}}{\mathrm{d}t}\vec{\Psi}_{\alpha x c}$ is calculated from Eq.\ \eqref{eq:PsiDef}. 
The efficiency of the present method comes from the absence of matrix-matrix products in Eqs. \eqref{eq:psiEOM} and \eqref{eq:omgEOM}. More significantly, Eqs. \eqref{eq:sigEOM}, \eqref{eq:psiEOM} and \eqref{eq:omgEOM} constitute a regular, densely coupled ODE, for which there are standard methods of solving, given initial conditions. This also implies  that a calculation can be restarted from the state obtained from previous time-steps. Importantly, it also scales linearly in the number of time-steps.

A scaling analysis of the method is furthermore given in Fig. 2 of Ref. \cite{Popescu2016}. \added[id=ABL4]{Details of the scaling can also be found in App. \ref{app:scaling}.} \deleted[id=ABL4]{The scaling is determined by the largest number ($n_o$) of eigenvalues of the coefficient matrices $\mathbf{W}_{\alpha l}$. }
\deleted[id=ABL4]{The scaling is linear in system size when $n_o  \approx \mathrm{dim}(\bsigma)$, and change to cubic scaling when $n_o \ll \mathrm{dim}(\bsigma)$. This scaling behavior is a result of the commutator in Eq. \eqref{eq:sigEOM} becoming the dominant factor in terms of computation time for large systems. In terms of number of poles the scaling is quadratic as can be seen from Eqs. \eqref{eq:psiEOM} and \eqref{eq:omgEOM}.}

For the steady-state, an equation can furthermore be derived by letting $\Delta_\alpha(t)=c_\alpha$ where $c_\alpha$ is a constant initial bias, and setting the time-derivatives of $\Vec{\Psi}_{\alpha x c}$ and $\Omega_{\alpha x c\alpha' x' c'}$ in Eqs.\ \eqref{eq:psiEOM} and \eqref{eq:omgEOM} to zero. Doing this, the following equations are obtained:
\begin{align}
    \label{eq:Omega0eq}
    \Omega_{\alpha x c \alpha' x' c'}^0&=-\Biggl((\Lambda^{>,-}_{\alpha' x' c'} - \Lambda^{<,-}_{\alpha' x' c'})\vec{\xi}^{\phantom{.}\dagger}_{\alpha' x' c'}\vec{\Psi}_{\alpha x c}^0\nonumber\\
    &\phantom{mn}+(\Lambda^{>,+}_{\alpha x c}\phantom{n.} - \phantom{n}\Lambda^{<,+}_{\alpha x c})\vec{\Psi^0}^\dagger_{\alpha' x' c'}\vec{\xi}_{\alpha x c}\Biggl)
    \\
    &\phantom{mn}\times \left[\frac{i\hbar}{\chi_{\alpha' x'}^-(t_0)- \chi_{\alpha x}^+(t_0)}\right]\nonumber
\end{align}
and
\begin{align}
    \label{eq:Psi0eq}
    \vec{\Psi}^0_{\alpha x c} &= -\left[ \mathbf{H}^0- \mathbf{1}\chi_{\alpha x c}^+  \right]^{-1}
    \\
    &\phantom{m} \times\Biggl(\hbar\Lambda_{\alpha x c}^{<+}\vec{\xi}_{\alpha x c}.\nonumber
    \\
    &\phantom{mm}+\hbar (\Lambda_{\alpha x c}^{>+} - \Lambda_{\alpha x c}^{<+})\bsigma^0\vec{\xi}_{\alpha x c}\nonumber \\
    &\phantom{mm}+\frac{1}{\hbar}\sum_{\alpha' x' c'}\Omega_{\alpha x c \alpha' x' c'}^0 \vec{\xi}_{\alpha' x' c'}\Biggl)\;.\nonumber
\end{align}
This system can be solved by substituting $\Omega_{\alpha x c \alpha' x' c'}^0$ into Eq.\ \eqref{eq:Psi0eq} and noting the equation is a linear inhomogeneous matrix problem. The system of equations just requires the steady-state density matrix to be known. Techniques for calculating the steady-state density matrix both in and out of equilibrium are well-developed\cite{Papior2017, stokbro2003transiesta, brandbyge2002density}. It is obtained by using the Keldysh equation for the lesser Green's function as \cite{haug2008quantum},
\begin{align}
    \label{eq:eq_DM}
    \bsigma^0 = 
    \frac{i}{2\pi} \int_{-\infty}^{+\infty} \mathbf{G}^r(\epsilon)\mathbf{\Sigma}^<(\epsilon)\mathbf{G}^a(\epsilon)\mathrm{d}\epsilon\,.
\end{align}
This integral can be handled using contour integration, and the energy-resolved self-energy from Eq. \eqref{eq:L_SE} in steady state. Furthermore, it simplifies with the fluctuation-dissipation theorem under equilibrium conditions \cite{haug2008quantum}. It should also be noted the static expression in Eq. \eqref{eq:eq_DM} also accounts for bound states as $G^{r/a}(\epsilon)$ contains all the the spectral information of the device. The same applies in the time-dependent case, where the bound states are likewise accounted for in equilibrium\cite{stefanucci2007bound}. Out of equilibrium needs a fully time-dependent treatment to determine the density matrix\cite{stefanucci2007bound}.

Lastly, it is also possible to introduce a general additive  term in Eq.\ \eqref{eq:sigEOM}\added[id=ABL3]{.} This could be a dissipative term scaling with the deviation of the density matrix from equilibrium such as \cite{boyd2020nonlinear},
\begin{equation}
    \label{eq:Dissipation}
    \boldsymbol{\mathcal{D}}(t) = \eta (\bsigma(t) - \bsigma^0)\;.
\end{equation}
This term can e.g. be used to dampen rapid oscillations when calculating the initial state or introduce an {\it ad hoc} lifetime of electrons in the device. It can also be used to introduce a dampening term similar to what is used in the DLvN approach to quantum transport\cite{ramirez2019driven}. 

Altogether, the user must (1) start from a standard transport calculation using, for example, TBtrans \cite{papior2017improvements}, (2) obtain a fitted level-width function in the form of Eq. \eqref{eq:GammaForm}, a Fermi function expansion of the form Eq.\ \eqref{eq:FermiFunc}, (3) compute Eq.\ \eqref{eq:eq_DM} self-consistently to then obtain the steady-state solution $\vec{\Psi}^0_{\alpha x c}$ and $ \Omega^0_{\alpha x c \alpha' x' c'}$, and then, finally, (4) propagate Eqs. \eqref{eq:sigEOM}, \eqref{eq:psiEOM} and \eqref{eq:omgEOM}. The implementation is done in a stepwise manner, where the users can refine the fit as they see fit.

\section{Orthogonalization Procedure}
\label{sec:orthogonalization}
In Section \ref{sec:theory}, an orthogonal basis was assumed throughout. It is important to address the translation of a general DFT-based calculation, where the basis is not orthogonal, to the theory presented above. The main problem is to include the basis set overlap between the electrode basis $\Phi_\alpha$, and the device basis $\Phi_D$. To resolve this issue, the device basis is transformed according to Ref.~\cite{kwok2013time},

\begin{align}
	\left[\begin{array}{c}{{\Phi_{L}}}\\ {{\tilde{\Phi}_{D}}}\\ {{\Phi_{R}}}\end{array}\right]=\left[\begin{array}{c c c}{{\mathbf{1}}}&{{0}}&{{0}}\\ {{\mathbf{U}_{D L}}}&{{\mathbf{U}_{D D}}}&{{\mathbf{U}_{D R}}}\\ {{0}}&{{0}}&{{\mathbf{1}}}\end{array}\right]\left[\begin{array}{c}{{\Phi_{L}}}\\ {{\Phi_{D}}}\\ {{\Phi_{R}}}\end{array}\right]\, ,
\end{align}
such that the resulting overlap matrix of the whole system in the transformed basis is on the form,
\begin{align}
\tilde{\mathbf{S}}=\mathbf{U} \mathbf{S} \mathbf{U}^{T}=\left[\begin{array}{c c c}{{\mathbf{S}_{L}}}&{{0}}&{{0}}\\ {{0}}&{{\tilde{\mathbf{S}}_{D}}}&{{0}}\\ {{0}}&{{0}}&{{\mathbf{S}_{R}}}\end{array}\right].
\end{align}
This ensures the electrode and device basis functions have zero overlap. The blocks $\mathbf{U}_{DL}, \mathbf{U}_{DD}$ and $\mathbf{U}_{DR}$ are solved for in Ref. \cite{kwok2013time} and applying the transformation $\mathbf{U}$ to the total system results in a modification of the Hamiltonian, overlap and self-energies,
\begin{align}
    \tilde{\mathbf{H}}(t) = \mathbf{H}^0 + \sum_{\alpha} \mathbf{\Sigma}^0_\alpha(t), \label{eq:H_tdcor} \\
    \tilde{\mathbf{S}} = \mathbf{S} - \sum_{\alpha} \mathbf{\Sigma}^1_\alpha, \label{eq:S_tdcor} \\
    \tilde{\mathbf{\Sigma}}^r_\alpha(z) = \mathbf{\Sigma}^r_\alpha(z) - \mathbf{\Sigma}^0_\alpha - \mathbf{\Sigma}^1_\alpha z.
\end{align}
The expressions for $\mathbf{\Sigma}_\alpha^{0/1}$ can be found in App.~\ref{app:sig01}. The time-dependent Hamiltonian $\tilde{\mathbf{H}}(t)$ contains the time-dependent correction introduced by the electrode-device overlap $\mathbf{S}_{\alpha D}$. Furthermore, as the transformed device basis, $\Phi_D$, is orthogonal to the electrode basis, the device basis can be simply orthogonalized with a Löwdin transformation:
\begin{align}
    \label{eq:lowdintransform}
    \mathbf{M}'_D = \tilde{\mathbf{S}}^{-\frac{1}{2}}\mathbf{M}_D\tilde{\mathbf{S}}^{-\frac{1}{2}}.
\end{align}
This procedure makes the non-orthogonal electrode-device transport calculation compatible with the theory outlined in Section ~\ref{sec:theory}.

\section{Time-dependent Hamiltonian}\label{sec:Linerization}
Involving some type of density matrix dependence in the Hamiltonian $\mathbf{H}(t) = \mathbf{H}\left[\bsigma(t)\right]$, as done in a mean-field description such as TDDFT, will incur a cost in terms of computation time when calculating the Hamiltonian from the density matrix. In some cases it can be advantageous to approximate the Hamiltonian's dependence on the density matrix, $\bsigma$. Since Coulomb screening will keep the device region close to charge neutral, a Taylor expansion of the $\bsigma$ dependence of the Hamiltonian around the equilibrium density can be a good approximation. In this work, the Hamiltonian will be linearized in terms of the density matrix by proxy, through the Mulliken charge matrix \cite{hourahine2020dftb+}
\begin{align}\label{eq:mulliken}
    \mathbf{Q}_0 =  \frac{1}{2}\left[\bsigma^{NO}\mathbf{S} + \mathbf{S}\,\bsigma^{NO} \right].
\end{align}
Here,, $Q_0$ is the Mulliken charge matrix, $\mathbf{S}$ is the overlapmatrix,x, and $\bsigma^{NO}$ is the density matrix in thnon-orthogonalnal basis. The diagonal of $Q_0$ is the collection of the orbital-resolved Mulliken charge. In order to Taylor-expand $\mathbf{H}$ around the ground state $\bsigma^{NO}$, a small charge $\mathrm{d}q$ is added on the diagonal to give:
\begin{align}
    [\mathbf{Q}_0^i]_{jk} = [Q_0]_{jk} + \delta_{jk}\delta_{ki}\mathrm{d}q, 
\end{align}
which now is a displaced Mulliken charge matrix. To obtain the corresponding displaced density matrix from the Mulliken charge matrix $Q_0^i$, a Sylvester equation
\begin{align}
    \mathbf{Q}_0^i = \frac{1}{2}\left[ \bsigma^{NO}_i \mathbf{S} +\mathbf{S} \,\bsigma^{NO}_i \right]
\end{align}

is solved for $\bsigma_i^{NO}$, from which $\mathbf{H}(\bsigma^{NO}_i)$ can be calculated using an external DFT-code. This in turn enables the calculation of the Taylor-expansion coefficients
\begin{align} 
    \frac{\mathrm{d}\mathbf{H}}{\mathrm{d}q_i} \approx \frac{\mathbf{H}(\bsigma_i^{NO}) - \mathbf{H}^0}{\mathrm{d} q}.
\end{align}
The linearized Hamiltonian dependence on the Mulliken charge will then read
\begin{align} \label{eq:taylor_exp}
    \mathbf{H}(\bsigma(t)) = \mathbf{H}^0 + \sum_i \frac{\mathrm{d}\mathbf{H}}{\mathrm{d}q_i}[\mathbf{Q}(\bsigma(t)) - \textbf{Q}_0]_{ii}
\end{align}
where the time-dependent Mulliken charge matrix will be calculated for each timestep. Other schemes, such as directly expanding the density matrix elements, are also possible. The main benefit is in any case, that the dynamic Hamiltonian from Eq.~\eqref{eq:taylor_exp}, can be evaluated fast compared to a full DFT calculation. 

Here, we treat the coupling to a classical photon field. However, it should be noted that in Ref. \cite{tuovinen2024electroluminescence} the coupling to photonic cavity quantum modes is incorporated into Eq. \eqref{eq:sigEOM} by extending the time-evolution to two sets of dynamical variables describing both electrons and modes, respectively.

\section{Examples}\label{sec:results}
In this Section, three different systems are examined under influence of a THz-pulse. The first is a tight-binding model incorporating a spin-splitting Hubbard $U$ term to model hydrogen adsorption on graphene. For this simple electronic structure model we also include
an additional infrared pump field. The second system is a model of a THz-STM measurement of an arm-chair nanoribbon in which the electronic structure is described in greater detail using the DFTB+ code. The third system is a gold break-junction and where the significance of the gap on the current-response is examined using a linearized Mulliken-charge dependence of the Hamiltonian obtained with the SIESTA code.

Throughout this Section, the adiabatic approximation for the exchange-correlation functional will be used\cite{ullrich2011time}, i.e., no frequency-dependent effects are included, meaning the Hamiltonian depends only on the instantaneous density $\bsigma(t)$. This is done, however, only for practical reasons, and a retarded density matrix dependence of the Hamiltonian could be implemented without any theoretical barriers.
\subsection{Tight-binding with Mean-field Hubbard Term: Hydrogenated Graphene} \label{sec:Res1_1}
Graphene can be hydrogenated with a hydrogen atom on top of a carbon atom, which induces a magnetic moment in the graphene layer \cite{gonzalez2016atomic}.
This has also been treated in a simplified form in Ref.\ \cite{svaneborg11manipulation}. The basic property of this system is a switchable spin-polarization of the hydrogen atom when contacted with a STM-tip from above. 

\subsubsection*{Model}
Here an extended graphene sheet hydrogenated at a single carbon atom is modeled. The general setup is illustrated in Fig.\ \ref{fig:H_gr_stm}.
\begin{figure}
    \centering
    \includegraphics[width=0.4\textwidth]{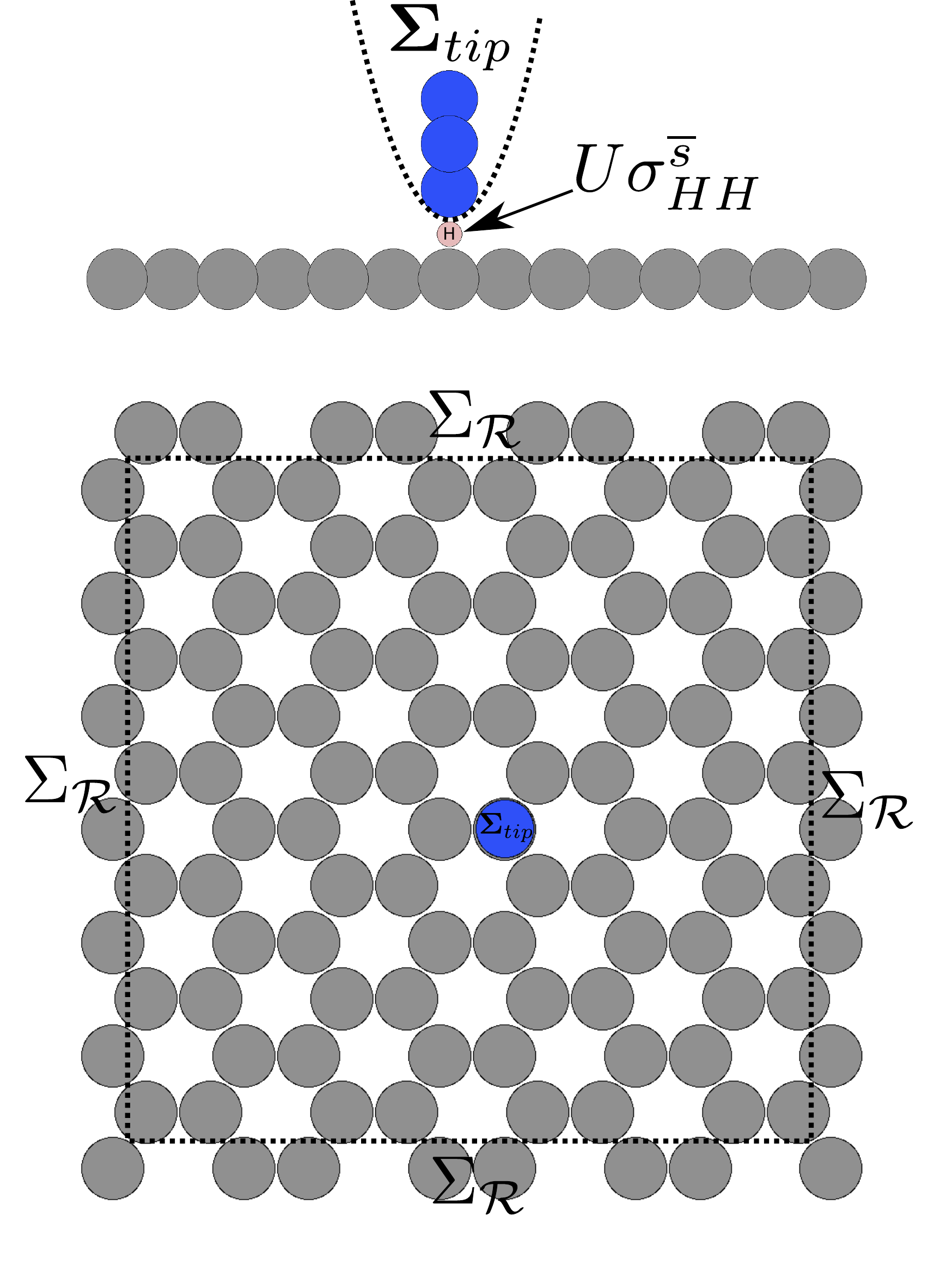}
    \caption{Computational setup for the hydrogen atom on graphene. One orbital for each atom is added, giving a total of $85$ orbitals.}
    \label{fig:H_gr_stm}
\end{figure}
A mean-field Hubbard term $U\bsigma_{HH}^{\overline{s}}$ is indicated where $\overline{s}$ means the opposite spin \cite{Sanz2022}. A single-band electrode self-energy $\mathbf{\Sigma}_{tip}$ and the graphene real-space self-energy $\mathbf{\Sigma}_{\mathcal{R}}$ are also shown \cite{papior2019removing}. The full Hamiltonian consisting of the various parts is given as
\begin{align}
    H_F = H_{gr} + H_{\rm tip} + H_{\rm dip} + H_H + \left[H_{ gr,H} + H_{H,\mathrm{tip}} + \mathrm{H.c}\right],
\end{align}
where the different terms are
\begin{align}
    \label{eq:R1_Ham}
    H_{gr}  &= \sum_{<i,j>,s} \gamma_g c^\dagger_{is} c_{js} + \sum_{is}V^i(t)c^\dagger_{is}c_{is}\,,\nonumber\\
    H_{tip} &= \sum_{<i,j>,s} \gamma_t t^\dagger_{is} t_{js} + \sum_{i,s}V_{tip}(t) t_{is}^\dagger t_{is}\,,\nonumber\\
    H_{H}   &= \sum_{s}(\epsilon_H + V_{tip}(t) + U\bsigma_{HH}^{\overline{s}}(t))d^\dagger_{Hs}d_{Hs}\,,\\
    H_{H,tip} &= \sum_{s}K_{tip}d^\dagger_{Hs} t_{0s}\nonumber\,,\\
    H_{H,gr} &= \sum_{s}K_{gr}d^\dagger_{Hs} c_{0s}\,,\nonumber\\
    H_{dip} &= \sum_{<i,j>,s} \Vec{r}_{ij}\cdot\Vec{E}(t) c_{is}^\dagger c_{js}\nonumber\,.
\end{align}
Here, $<i,j>$ means nearest neighbors, $s$ is a spin-index and $c_{is}$, $t_{is}$ and $d_{is}$ are the annihilation operators on each site. The parameters taken in the Hamiltonian are similar to the ones in Ref.\cite{svaneborg11manipulation}, namely $\gamma_g=-2.7$eV, $\gamma_t=-5.0$eV, $\epsilon_H=-1.7$eV, $U=6.5$eV, $K_{tip}=0.25$eV, $K_{gr}=3.25$eV , $\mu_{gr}=\mu_{tip}=0$eV and $kT_{gr} =kT_{tip}= 0.025$eV. The time-dependent bias $V_{tip}(t)$ is taken to be the experimentally measured THz-pulse seen in Fig.\ \ref{fig:AP_fig}.  In this model, the hydrogen on-site energy follows the tip chemical potential ($\mu_{tip}(t)= \mathrm{e}V_{tip}(t)$), while the graphene chemical potential $\mu_{gr}=0$ is kept at zero. The dipole Hamiltonian $H_{dip}$ includes a rapidly varying, spatially constant electric field in the dipole approximation. The real-space self-energy is in this case calculated using the sisl \cite{zerothi_sisl, papior2019removing} and the STM-tip self-energy is calculated using TBtrans with the DFT and transport code siesta\_python \cite{papior2017improvements, sancho1985highly}. Any field-dependent coupling between the hydrogen and tip are neglected, meaning $K_{tip}$ in eq. \eqref{eq:R1_Ham} is taken as constant.

\begin{figure*}
    \centering
    \includegraphics[width=0.7\textwidth]{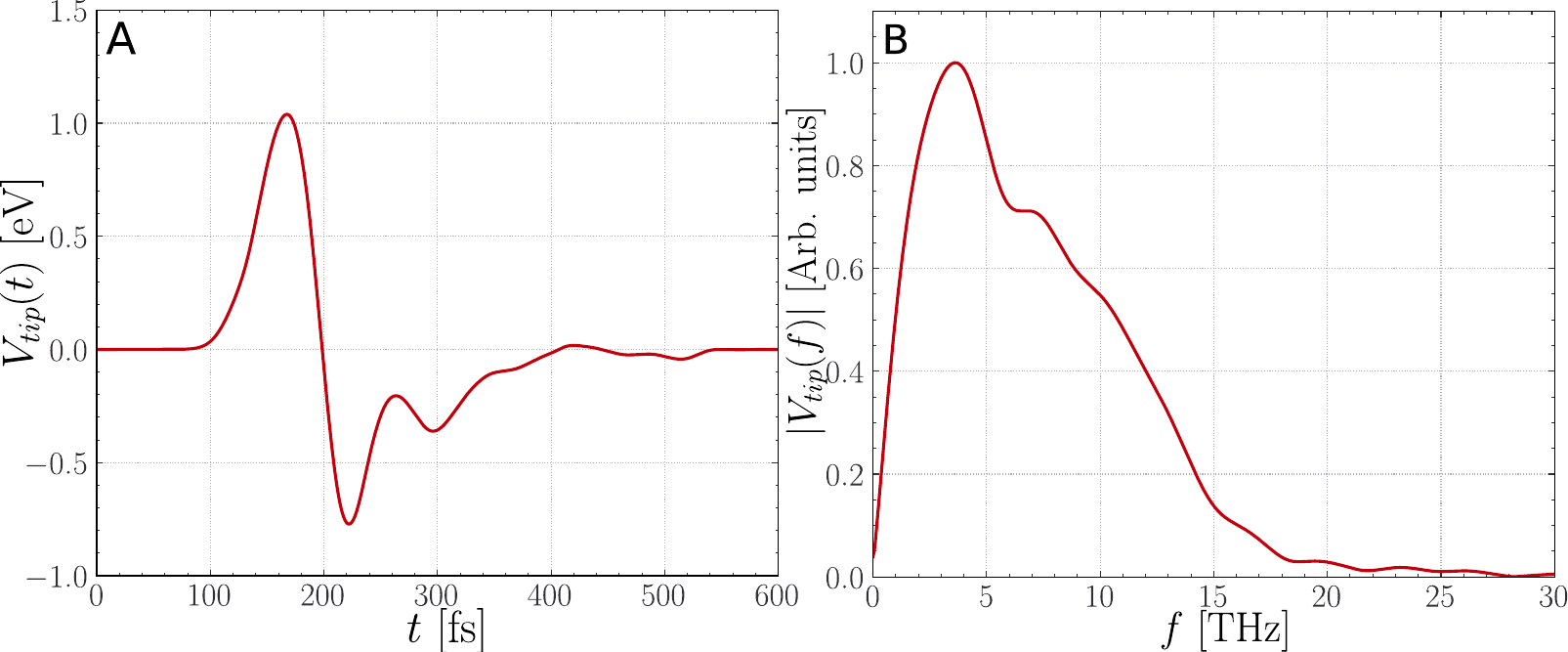}
    \caption{Temporal shape and frequency content of the experimental THz pulse $V_{tip}(t)$ used in the calculations.}
    \label{fig:AP_fig}
\end{figure*}

\begin{figure*}
    \centering
    \includegraphics[width=0.7\textwidth]{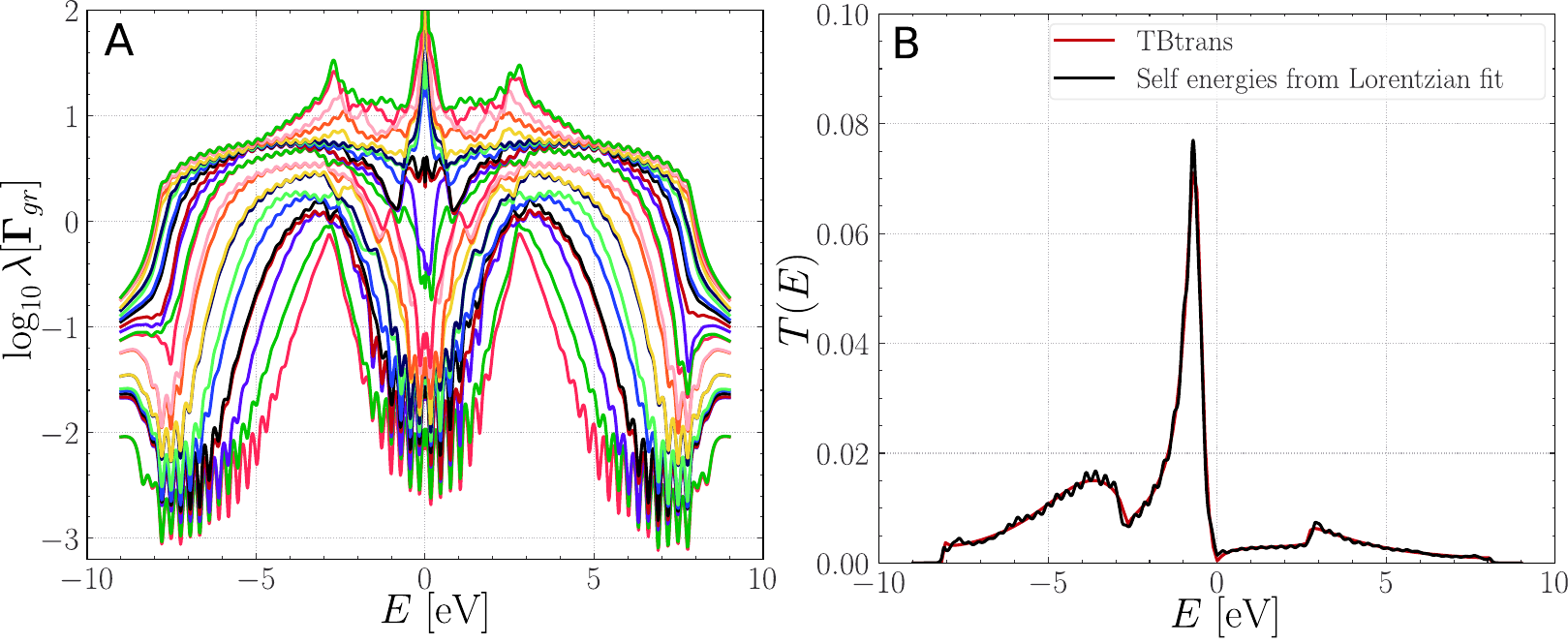}
    \caption{A) Eigenvalues of the fitted level-width function $\mathbf{\Gamma}_{gr}$ as a function of energy. B) Transmission function with exact self-energies and self-energies obtained from the fit, calculated without the Hubbard term.}
    \label{fig:gamma_gr_eig}
\end{figure*}

The eigenvalues of the fitted $\mathbf{\Gamma}_{gr}(\epsilon)$ can be seen in Fig.\ \ref{fig:gamma_gr_eig} while the single-site $\mathbf{\Gamma}_{tip}$ is easily verified to be positive semi-definite. Furthermore, note the significant pole in $\mathbf{\Gamma}_{gr}$ at $E=0$eV. 

\subsubsection*{Initial State}
The ODE to be propagated is now defined with the device part of the Hamiltonian from Eq.\ \eqref{eq:R1_Ham} and the pulse in Fig.\ \ref{fig:AP_fig}. The last ingredient to consider is initial steady-state density matrix for $t\rightarrow -\infty$ which in turn determines the auxiliary mode vectors through Eq.\ \eqref{eq:Psi0eq}. This means solving Eqs.\ \eqref{eq:eq_DM} and \eqref{eq:R1_Ham} self-consistently for the steady-state solution $\bsigma^0_{s}(t=-\infty)$. This is done using the {\tt SCF} tool of the Zandpack package. There are two possible solutions, namely a spin-unpolarized hydrogen orbital and a spin-polarized hydrogen orbital \cite{svaneborg11manipulation}. However, if the system is started in the unpolarized state, it will decay into the polarized state. Furthermore, the difference between starting the calculation with the state obtained using Zandpack's {\tt SCF} and {\tt psinought} tools (SCF-$\Psi^0$ initial state) and the naive approach $\Vec{\Psi}_{\alpha x c}(t_0)=\Vec{0}$ and $\Omega_{\alpha x c\alpha' x' c'}(t_0)=0$ can be seen in Fig.\ \ref{fig:ISdiff}, where both starting states have been propagated using Zandpack's {\tt zand} tool.  The SCF-$\Psi^0$ initial state is clearly seen being very close to the actual steady state, while the naive starting state has large oscillations before approaching the steady state.
\begin{figure}
    \centering
    \includegraphics[width=0.4\textwidth]{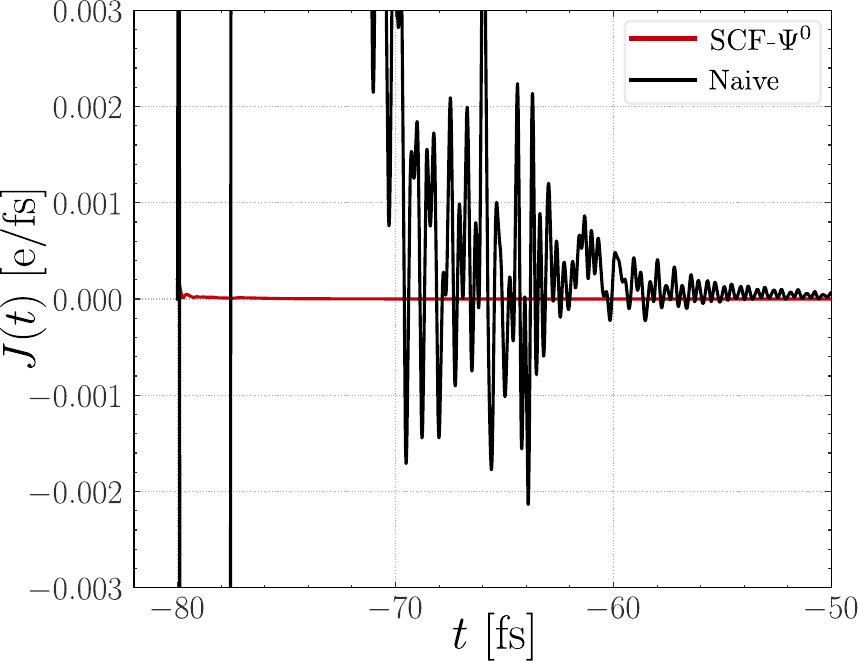}
    \caption{The current through the device using the SCF-$\Psi^0$ initial state and compared to using the naive initial state, and same initial density matrix.  The near-vertical black lines at $-80$fs$ < t < -78$fs are indicating the rapidly varying current.}
    \label{fig:ISdiff}
\end{figure}
The SCF-$\Psi^0$ initial state furthermore enables the adaptive integrator implemented to solve the EOM to take significantly larger steps in the beginning, lowering the time spent on the initial times before the pulse perturbs the system. The SCF-$\Psi^0$ is however not completely steady-state, but still has a small but noticeable current  close to $t=-80$fs in Fig.\ \ref{fig:ISdiff}. 
The reason for this is the {\tt SCF} tool determines $\bsigma^0_s$ with an exact Fermi-function and a contour whose lower bound and sampling density has to be converged, while in the time-propagation, the Fermi-function is approximated by Eq.\ \eqref{eq:FermiFunc} and the contour integral is exact. Therefore comparing the two calculations that use different contours for determining the steady state allows one to determine if the number of poles in the Fermi-function expansion is sufficient.
\begin{figure*}
    \centering
    \includegraphics[width=0.95\linewidth]{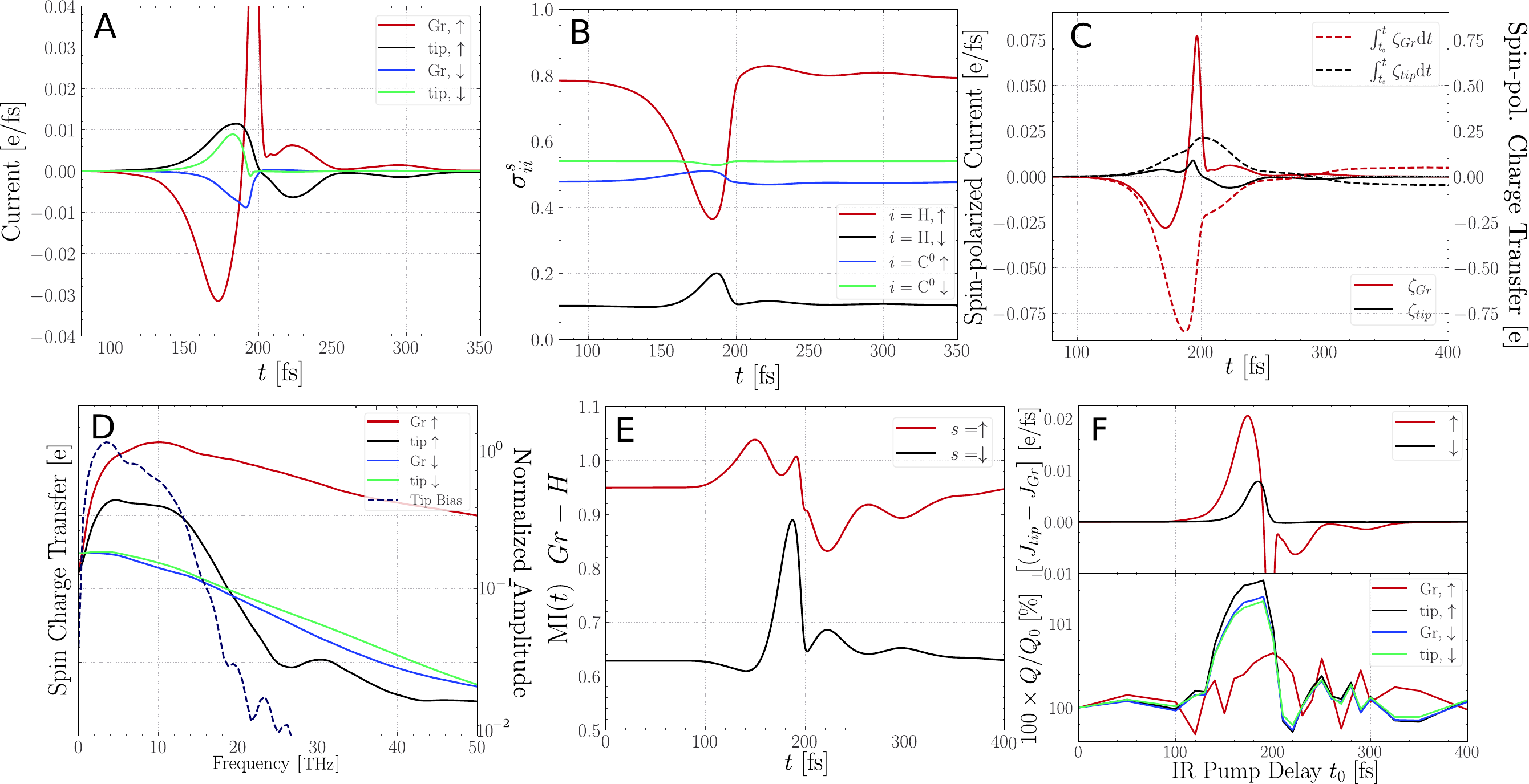}
    \caption{A) Spin-polarised currents in each electrode. B) Occupation on hydrogen orbital and carbon orbital sitting below the hydrogen\added[id=ABL3]{ ($\mathrm{C}^0$)}. C) Spin-polarized current and spin-charge transfer. D) Fourier amplitude spectrum of the currents, together with the frequency-spectrum of the driving THz pulse. The amplitude spectrum of the currents have been normalised so their relative sizes are correct. E) Mutual information between hydrogen orbital and graphene sheet. F) Bottom panel shows the change in charge transfer in $\%$ as a function of the infrared pump delay time, while the top panel shows the current through the device (without IR pump) at various times for comparison.}
    \label{fig:R1Fig1}
\end{figure*}

\subsubsection*{Time-dependent NEGF Calculations}
A time-dependent bias on the hydrogen orbital is now included and Eqs. \eqref{eq:sigEOM}, \eqref{eq:psiEOM} and \eqref{eq:omgEOM} are propagated using Zandpack's {\tt zand} tool.
The result without dipole interaction ($H_{\rm dip} = 0$ in Eq.\ \eqref{eq:R1_Ham}) can be seen in Fig.\ \ref{fig:R1Fig1}A and shows a sharp rise in current at the moment the system decays from a spin-polarized state to a spin-unpolarized state. The actual transition between these two states can be seen in Fig.\ \ref{fig:R1Fig1}B where the spin up and down occupation on the hydrogen site is seen to suddenly approach each other. The same is the case for the carbon-atom sitting under the hydrogen atom to a lesser extent. In Fig.\ \ref{fig:R1Fig1}C the spin-polarised current $\zeta_\alpha(t) =J^\uparrow_\alpha(t) -J^\downarrow_\alpha(t)$ is plotted together the time-dependent spin-charge transfer. It shows a large transfer of spin during the peak of the pulse, which is subsequently reversed by the spin-polarized to spin-unpolarized transition. The frequency spectrum in Fig. \ref{fig:R1Fig1}D furthermore shows that there is a large generation of current at other frequencies than what is present in the pulse. This comes from the non-linear current-voltage relationship that gives rise to the current not just being a scaled copy of the applied bias. Furthermore, there is a significant difference in the rectified current for the spin-up and spin-down channels, owing to the difference in current response for each spin channel, as seen Fig. \ref{fig:R1Fig1}A. Lastly, the correlation between the graphene and the hydrogen orbital can be quantified through the mutual information \cite{henderson2001classical, sharma2015landauer, bergmann2021green}
\begin{align}
    \mathrm{MI}_{AB} = S([\bsigma]_A)+S([\bsigma]_B) - S([\bsigma]_{A\cup B}),
\end{align}
where the brackets $[ \bullet ]_K$ mean the projection on subsystem $K$ and $S(\bsigma) = -\mathrm{Tr}\left[\bsigma \log\bsigma\right] - \mathrm{Tr}\left[(1-\bsigma) \log(1-\bsigma)\right]$ is the entropy. The mutual information for each spin can be seen in Fig.\ \ref{fig:R1Fig1}E and shows a notable increase in correlation for the spin-down case as the pulse passes through the system.

Next, a weak spatially uniform infrared linearly polarized electric field of the form
$\Vec{E}(t)= \hat{\Vec{x}}E(t)$
is included. It has  wavelength $\lambda=1035$nm and interacts through a dipole interaction 
\begin{align}
    H_{dip} = V_{dip}(t)\sum_{<i,j>, s}\frac{x_{ij}}{|\Vec{r}_{ij}|}c_{is}^\dagger c_{js}\,,
\end{align}
where 
\begin{align}
    V_{dip}(t)=0.01\mathrm{eV} \cdot f(t)\,.
\end{align}
The pulse shape $f$ of the infrared field is taken as 
\begin{align}
f(t) = \mathrm{e}^{-\left[\frac{t - t_0}{s_0}\right]^2}\cos(2\pi f_0 (t - t_0))\,,
\end{align}
where $s_0 = 15$fs and $f_0 = \frac{v_{light}}{\lambdabar}\approx 0.29\mathrm{fs}^{-1}$, with $v_{light}$ the speed of light and $\lambdabar$ the wavelength. $t_0$ is the variable relative delay of the IR pump pulse to the THz probe pulse. The total charge transfer from the pulse $Q_\alpha^s = \int_{-\infty}^\infty J_\alpha^s(t)\mathrm{d}t$ can then be plotted as a function of the IR pump delay $t_0$. This is seen in Fig.\ \ref{fig:R1Fig1}F and shows a prominent peak in rectified charge at $t_0=195$fs, nicely coinciding with the point of the transition from spin-polarized to unpolarized from Fig.\ \ref{fig:R1Fig1}B.

In an actual experiment the situation modeled here is repeated with MHz frequency, meaning the charge-transfer $Q_\alpha^s$ in Fig.\ \ref{fig:R1Fig1}F manifests as a change in measured current in the tip. The relative pump-probe delay $t_0$ can furthermore be accurately controlled, contrary to the absolute time of the peak in $V_{tip}$\cite{yoshida2021terahertz, garg2022real, jelic2024atomic}.
\subsection{DFTB+ Interface: STM Tip on Armchair Graphene Nanoribbon}\label{sec:Res2}
In Ref.\ \cite{ammerman2021lightwave}, experimental measurements have been reported for AGNR nanoribbons which show a rectified current in a THz-STM setup. Here this system is simulated using the DFTB+ code\cite{aradi2007dftb+} to give the Hamiltonian as the density evolves with time. {The only complication in terms of theory is that the basis used  by LCAO-based DFT-codes are not generally orthogonal, but the basis in which the EOM are formulated is orthogonal.} It is, however, straightforward to translate between the orthogonal and non-orthogonal bases when using a Löwdin transformation to orthogonalize the states and the correction terms mentioned in Section ~\ref{sec:orthogonalization}. The calculations runs along the same vein as in subsection \ref{sec:Res1_1}. The SCF-$\Psi^0$ solution obtained with the {\tt SCF} and {\tt psinought} tools as before. DFTB+ calculates the Hamiltonian in terms of the Mulliken charges, $q_i$, on each orbital, $i$. The recipe for transformation from the orthogonal basis density matrix to the Mulliken charges, $q_i$, is done by calculating the non-orthogonal density, matrix\cite{hourahine2020dftb+}
\begin{align}
    \label{eq:MullCharge_2}
    \sigma^{NO} = \mathbf{S}^{-1/2}\bsigma^O \mathbf{S}^{-1/2},
\end{align}
where $\mathbf{S}$ is the overlap matrix and then use Eq. \eqref{eq:mulliken}. The process of obtaining the $\mathbf{\Gamma}_\alpha$'s of the electrodes is then the same as in the TranSIESTA method, where the pristine electrode Hamiltonian is calculated, now using the DFTB+ code. The interface is implemented in the siesta\_python code \cite{siesta_python_web}.

The device structure can be seen in Fig.\ \ref{fig:AGNR_setup}A and \ref{fig:AGNR_setup}B and the electrode electronic structure can be seen in Fig.\ \ref{fig:AGNR_setup}C and \ref{fig:AGNR_setup}D. The fit of $\mathbf{\Gamma}_\alpha$ is done the same way as for the hydrogenated graphene sheet, where the fitting scheme here also uses $N_l = 81$ Lorentzians for all three electrodes.  The device region furthermore contains $ 4\cdot N_{Al} + 1\cdot N_C = 4\cdot 16 + 1\cdot 224= 288$ orbitals active in the calculation. Only the $p_z$ part of the carbon atoms are used and the hydrogen $s$ orbitals are discarded, while all four $sp^3$ orbitals on the aluminium atoms are used.  
\begin{figure*}
    \centering
    \includegraphics[width=0.66\linewidth]{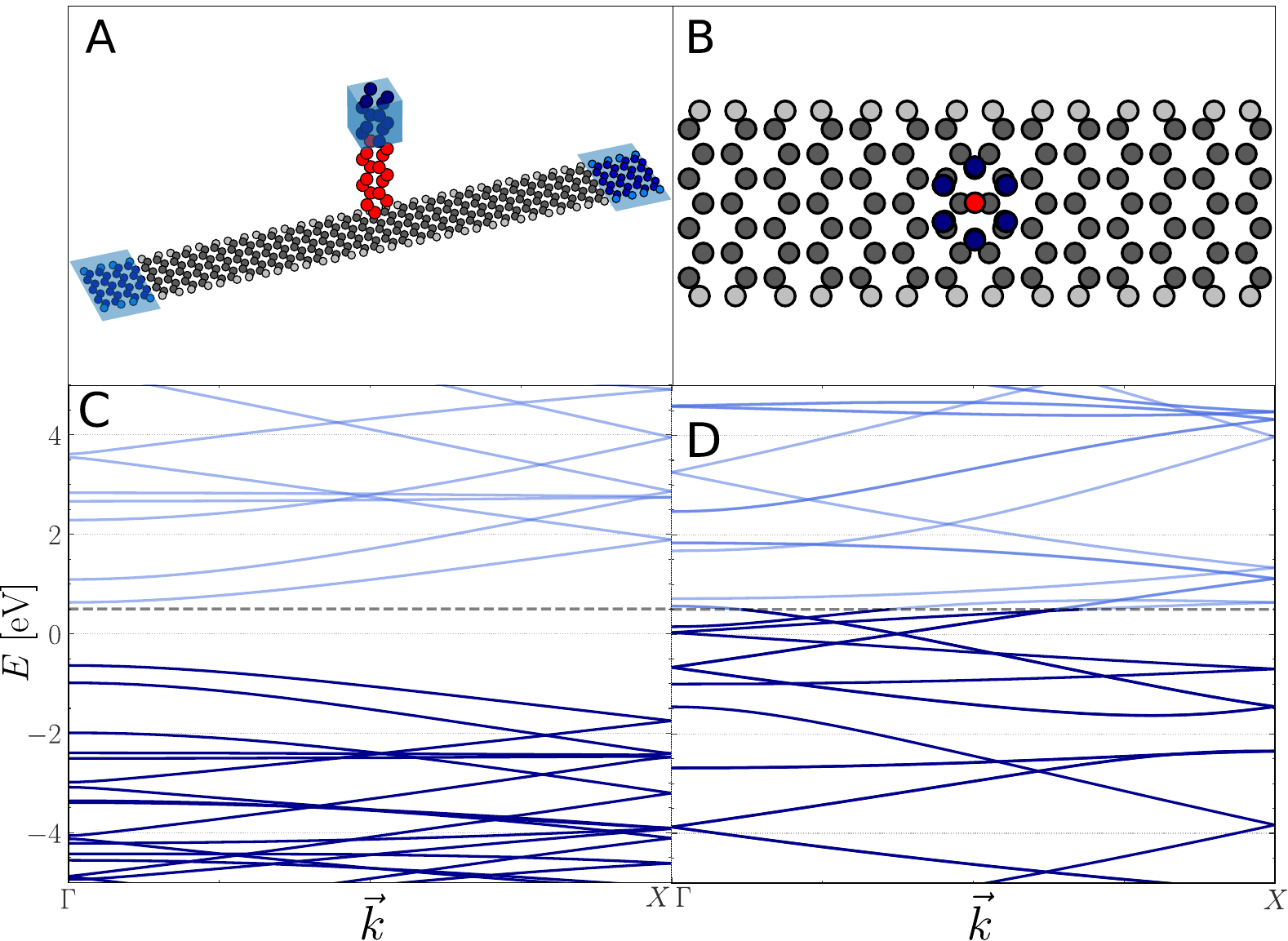}
    \caption{The AGNR with tip illustrated with associated electronic band structure. Sideview in A) and top view in B). Hydrogen in light grey, carbon in dark grey, aluminium in red, electrode atoms in shaded blue. C) Semi-conducting AGNR. D) Metallic Al Tip. Occupied states in dark blue, unoccupied states in light-blue. In C and D, the Fermi-level at equilibrium is taken at $E_F=0.5$eV, indicated by the two dashed lines.}
    \label{fig:AGNR_setup}
\end{figure*}

The AGNR is semi-conducting which is evident from the band-structure, while the aluminium tip is metallic. The goal is to model what happens as a strong THz pulse is incident on this system, causing what in steady-state would be a potential energy shift in the tip. The system parameters being time-dependent does not, strictly speaking, allow for a description in energy, unless the system is evolving adiabatically. The bands in Fig.\ \ref{fig:AGNR_setup}C,D are however still useful to think of in relation to THz frequency perturbations of the system. The pulse seen in Fig.\ \ref{fig:AP_fig} is applied as the tip bias
\begin{align}
    \Delta_{tip}(t) = V_{tip}(t)\,, 
\end{align}
while keeping the AGNR electrodes fixed at $\Delta_{L/R} = 0$. Before doing so, all chemical potentials are set to $\mu_\alpha = 0.5$eV relative to the pristine $\mu_\alpha = 0$eV. A ramp $V_{ramp}(z)$ that goes from 0 in the AGNR ($z=0$) to $\Delta_{tip}(t)$ in the tip electrode ($z=z_{max}$) is included, together with the density dependence of the Hamiltonian obtained from DFTB+. The current and rectified charge as the system evolves in time are seen in Fig.\ \ref{fig:AGNR_calculation_1}A together with the device charge in Fig.\ \ref{fig:AGNR_calculation_1}B. The current shows initially a normal scaling as the bias on the tip grows positive, but when the bias pulse instead becomes negative, rapid oscillations are seen in all three electrode currents. 
This behavior can be understood in the following way: When the tip electrode potential(bias) is positive(negative), tip electrons can tunnel from the tip into the unoccupied part of the spectrum seen in Fig.\ \ref{fig:AGNR_setup} of the AGNR resulting in a fairly smooth current curve in this part of the simulation. However, when trying to drive electrons from the semi-conducting ribbon to the tip, there are instead rapid oscillations, as seen in Fig.\ \ref{fig:AGNR_calculation_1}A when $V_{tip}<0$. 
The reason for these oscillations is that when trying to drive electrons from the occupied part of the AGNR into the tip, whose quasi-Fermi-level now is in the band-gap, the electrons can only do so by obtaining energy from a high-frequency oscillation as seen in Fig.\ \ref{fig:AGNR_calculation_1}A.
These oscillations subside again after the pulse is over. Furthermore, there is a large amount of charge-transfer relative to what was seen in Fig.\ \ref{fig:R1Fig1}C.

It is unknown if the current in the experiment of Ref.\ \cite{ammerman2021lightwave} is similar to the one in Fig.\ \ref{fig:AGNR_calculation_1}A since it is only the value of the rectified current, $Q_{tip}$, after the pulse has passed, which is measured. The $Q_{tip}$ does not show the rapid oscillations of the current, as it is the integral of it. In the STM measurement of Ref.\ \cite{ammerman2021lightwave}, metallic gold surface states introduced in the AGNR are furthermore observed, meaning the band-structure from Fig.\ \ref{fig:AGNR_setup} is not accurate for the electronic structure of the AGNR when it is modified by the substrate. They do, however, measure a rectified current corresponding to tens of electrons\cite{ammerman2021lightwave}, while for the pulse in Fig.\ \ref{fig:AP_fig} and the AGNR in Fig.\ \ref{fig:AGNR_setup}, around 0.7 electrons are rectified per pulse. One reason for this discrepancy might be the longer pulse duration in the experiment. Other factors are a stronger peak bias or contributions from the gold substrate.

The importance of the feedback of the density to change the Hamiltonian is seen when comparing the dynamic calculation from Fig.\ \ref{fig:AGNR_calculation_1}A to the one with a static Hamiltonian. The calculation with a static Hamiltonian can be seen in Fig.\ \ref{fig:AGNR_calculation_1}D, which displays a much larger current with none of the prominent high-frequency oscillations in Fig. \ref{fig:AGNR_calculation_1}A.
\begin{figure*}
    \centering
    \includegraphics[width=1.0\linewidth]{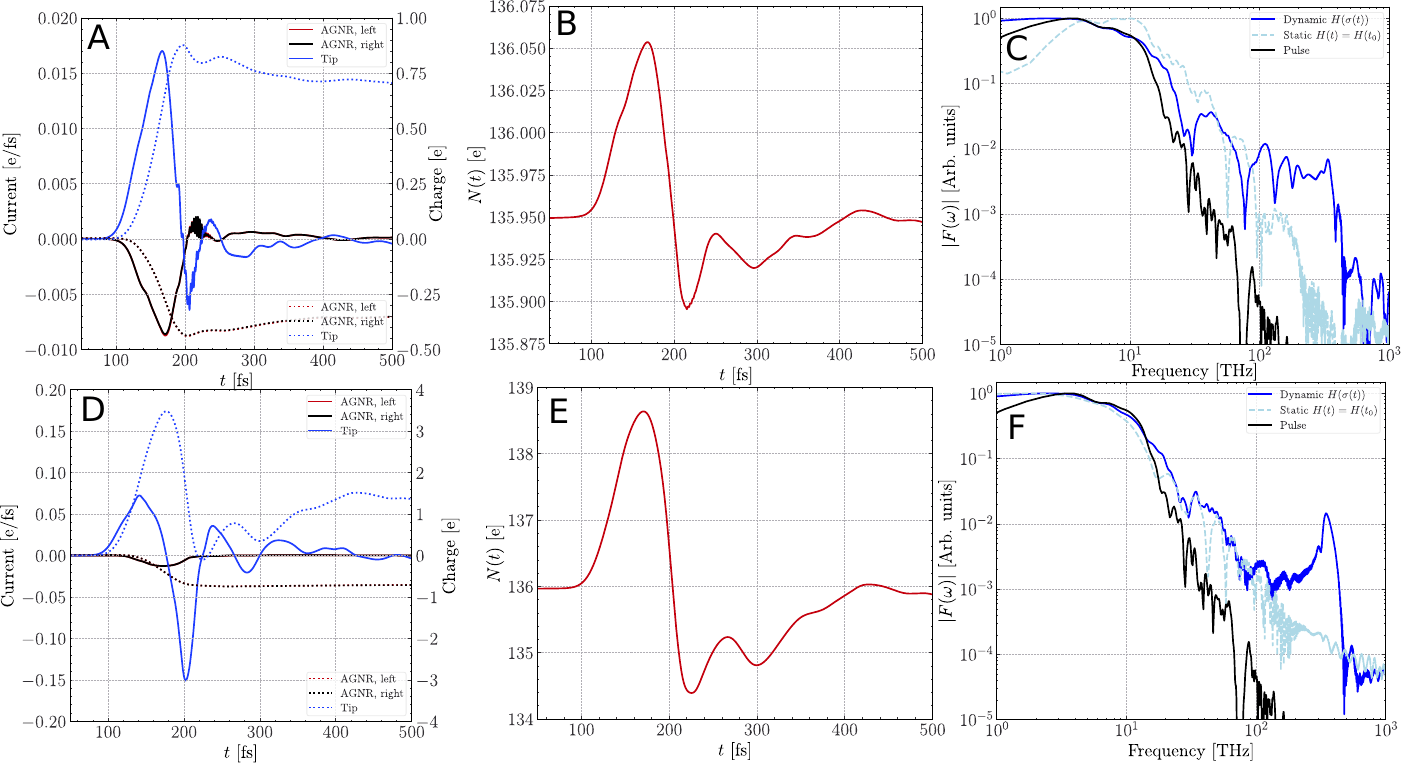}
    \caption{A) Current (full lines) and rectified charge (dotted lines) in each electrode as a function of time.  B) Device charge over time. C) Comparison of the tip current frequency content, calculated with and without the density dependence in the Hamiltonian.  D, E) Same calculation as in A and B, but with no density dependence in the Hamiltonian. F) Comparison of the AGNR current frequency content, calculated with and without the density dependence in the Hamiltonian.}
    \label{fig:AGNR_calculation_1}
\end{figure*}
The occupation number of the device region is also much larger in Fig.\ \ref{fig:AGNR_calculation_1}E compared to Fig.\ \ref{fig:AGNR_calculation_1}B because the electrons are not subjected to the electrostatic repulsion from the electronic density. The comparison of the frequency spectra of the currents can be seen in Figs.\ \ref{fig:AGNR_calculation_1}C and \ref{fig:AGNR_calculation_1}F and displays a much larger current response around $f = 350$THz corresponding to $ \sim 1.4$eV and is a result of an rapidly oscillating electron cloud that allows transition from valence band of the AGNR to the tip.

Furthermore, the frequency spectrum of the device charge can be obtained by tracing Eq.\ \eqref{eq:sigEOM} to obtain the continuity equation,
\begin{align}
    \frac{\mathrm{d}N}{\mathrm{d}t} = \sum_\alpha J_\alpha(t),
\end{align}
which in its Fourier-transformed version states $N(\omega) \propto \frac{1}{\omega} \sum_\alpha J_\alpha(\omega)$. 
\begin{figure}
    \centering
    \includegraphics[width=0.75\linewidth]{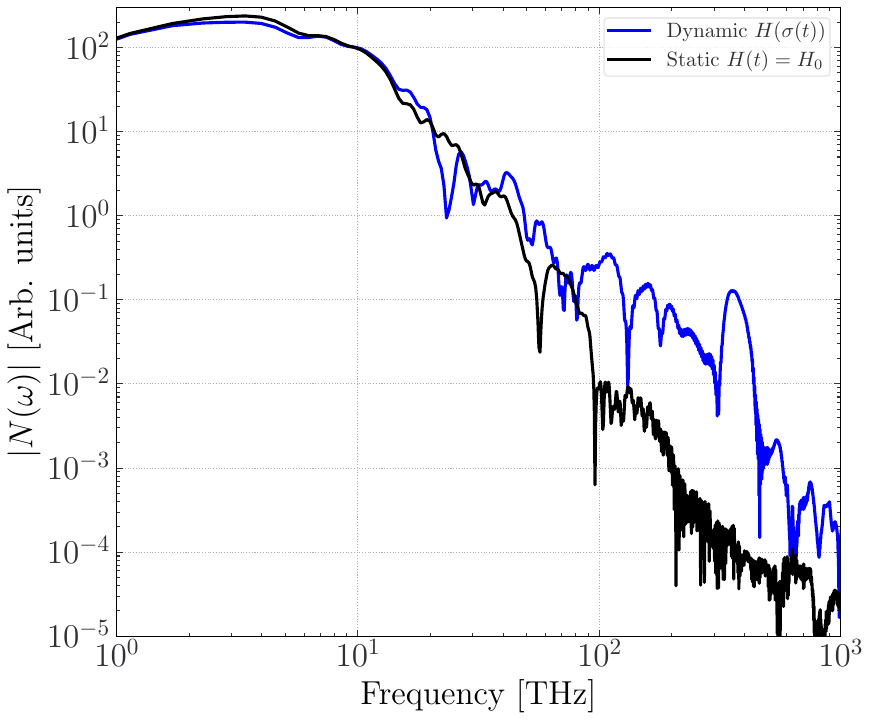}
    \caption{Normalized frequency spectrum of the device charge from the dynamical and static calculations from Fig.~\ref{fig:AGNR_calculation_1}. }
    \label{fig:AGNRFig_6}
\end{figure}

The Fourier transform of the device charge $N(\omega)$ can furthermore be taken as a measure of how large a dipole the system contains at frequency $\omega$. In Fig.\ \ref{fig:AGNRFig_6} we plot $N(\omega)$, which displays a small but significant dipole oscillation at $f=350$THz with an amplitude of about $\sim$1/100 compared to the maximum value of $N(\omega)$ located at $f\approx 4$THz. 

The radiated power, $P$, from an ideal, oscillating dipole is given as $P=\frac{\mu_0 \omega^4 p^2}{12\pi v_{light}}$ where $\mu_0$ is the permeability of free space, $\omega$ the oscillation frequency of the dipole and $p$ the dipole moment \cite{griffiths2005introduction}. Comparing now two dipoles, one oscillating with frequency $\omega_0$ and $p_0$, and a second oscillating with frequency $\omega_1 = 80\omega_0$ and $p_1 = p_0/100$, the second would then radiate $\frac{P_1}{P_0}\approx$ 4000 times more power than the first. It is however to be expected that dynamical screening effects\cite{hedin1999correlation, ridley2022many, ullrich2011time} should be included to account for the electronic response to density oscillations in this high-frequency regime. Such physics can be incorporated in the generalized Kadanoff-Baym ansatz (GKBA) methods such as developed in Refs.\ \cite{ridley2022many,tuovinen2020comparing}, but has not been incorporated in the method used in this work. It should also be stressed that it is an assumption that $N(\omega)$ is proportional to the actual dipole moment.

\begin{figure*}[h]
    \centering
    \includegraphics[width=0.75\linewidth]{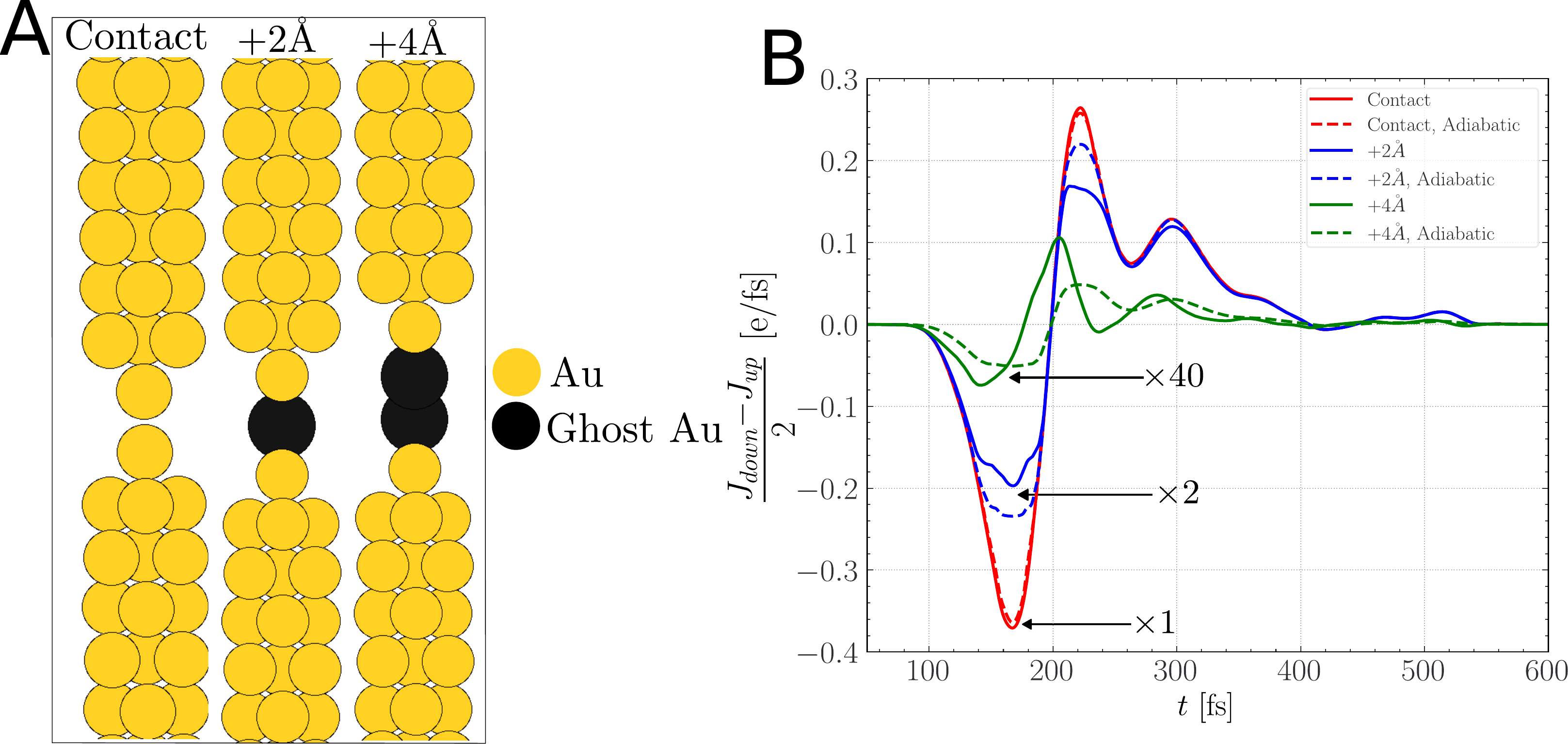}
    \caption{A) Gold tip geometries. B) Total currents through the geometries shown in A. Adiabatic results are also given for comparison. For each case, both the adiabatic and time-dependent current has been multiplied by the indicated factor. }
    \label{fig:GoldJunc_geom_curr}
\end{figure*}

\subsection{Gold junction}

The accuracy-level of DFTB+, which was used in Subsection \ref{sec:Res2}, is the so-called DFTB1 level \cite{hourahine2020dftb+}, which is characterized by the fact that the $\bsigma$-dependence of $\mathbf{H}$ is already linearized through the Mulliken charge given in Eq. \eqref{eq:mulliken}. This simple and  efficient approximation can also be useful for time-dependent calculations with a regular DFT code such as SIESTA instead of using the full dependence $\mathbf{H}(\bsigma)$. Here, the steady-state density matrix can be found, where after the Hamiltonian can be Taylor-expanded around the equilibrium Hamiltonian using a finite-difference approach, as described in Sec. \ref{sec:Linerization}.
We use the SIESTA code with the linearization to consider a gold atomic junction under the THz pulse.

The geometries under consideration for this Section are shown in Fig.\ \ref{fig:GoldJunc_geom_curr}A and indicated in black are ghost atoms which are used in the calculation. These introduce the same basis functions in the gap as are located on the gold atoms, but without the gold atom potential and electronic contribution. This in turn means that when the two gold tips are taken apart, the coupling does not artificially weaken due to basis function decay \cite{garcia2015ab, garcia2018tunneling}. 

For the calculation at hand, in addition to the linearization of the dependence of Hamiltonian on the density matrix, the system has an external potential induced by the THz field. It is introduced in the symmetric junction as a linear ramp in the $z$-direction with $V_{ext}(z) = 0$ at the centre of the junction and scaling with the $z$-component and bias $V_{bias}(t) = \Delta_1(t) - \Delta_0(t)$ such that the ramp matches the bias applied in each end of the junction. This potential is assumed only to be added on the diagonal of the non-orthogonal Hamiltonian through the potential
\begin{align}
    V_{ii}(t) = \frac{z_i - z_{centre}}{ z_{max} -  z_{min} }V_{bias}(t),
\end{align}
where $z_i$ is the $z$-component of the atom on which the orbital is located and $z_{centre}$, and $z_{min}$, $z_{max}$ the positions of the left and right electrodes, respectively. This potential is then readily brought into the orthogonal basis with the Löwdin transformation from Eq. \eqref{eq:lowdintransform}. The currents, as calculated using the previously outlined approach, are shown in Fig.\ \ref{fig:GoldJunc_geom_curr}B.
The adiabatic results are obtained using the Landauer-Büttiker formula \cite{Papior2017ImprovementsTransiesta} for the current in a non-interacting transport setup, using the instantaneous Hamiltonian $\mathbf{H}(t) = \mathbf{H}(\sigma(t))$ obtained from the fully time-dependent calculation. Significant differences between the adiabatic and time-dependent current are seen in the $+4$Å case where the coupling between the two tips is low. 
This example illustrates the importance of time-dependent method for sub-picosecond timescales.
\section{Conclusions}

In summary, we have refined the method proposed in Ref.\ \cite{Popescu2016} to be applicable to device sizes of experimental relevance and to use arbitrary electrodes. Additionally, we present an approach to solve the well-established steady-state NEGF equations within the framework of the auxiliary-mode method, which leads to a steady-state auxiliary-mode expansion and can be reduced to solving a system of equations. Using a novel decomposition of the line-width function established in Eq. \eqref{eq:GammaDecomp}, we can propagate the system within the time-dependent NEGF framework without obtaining spurious currents under equilibrium conditions, while also maintaining the efficiency of the eigendecomposition proposed in Ref. \cite{Popescu2016}.

The new generation of the auxiliary-mode NEGF method has been implemented in the Zandpack software package. It offers the possibility to include the complex density matrix dependence of the Hamiltonian through a general Python interface that can incorporate existing external DFT codes. Firstly, we demonstrate the use of Zandpack using a simple mean-field Hubbard model of hydrogenated graphene. We show calculations of pump-probe DC currents, which is usually measured in experiments, and of thermodynamical quantities such as entropy and mutual information. Secondly, we consider a scenario where the Hamiltonian is obtained by the DFTB+ package\cite{aradi2007dftb+, hourahine2020dftb+} and model a STM tip on an armchair graphene nanoribbon. The main bottleneck is typically the external code.

We have also demonstrated the linearization scheme based on the SIESTA code\cite{siesta,papior2017improvements}, and used the approximate density-matrix dependence of the Hamiltonian in the time-propagation. This linearization scheme is presented as a simpler solution that should still incorporate the main features of the density matrix dependence of the Hamiltonian from DFT.

Overall, Zandpack provides a versatile tool for time-dependent nanoelectronics simulations with interfaces to widely used tools. It is another step towards studying realistic and experimentally relevant scenarios.

\section{Funding Statement}
The Independent Research
Fund Denmark, grant number ”0135-00372A” has funded this work together with TED2021-132388B-C44 funded
by MCIN/AEI/10.13039/501100011033 and Unión Europea Next
Generation EU/PRTR, and grant PID2022-140845OB-C66 funded by
MCIN/AEI/10.13039/501100011033 and FEDER Una manera de hacer
Europa. We also acknowledge support from the Novo Nordisk Foundation Data Science Research Infrastructure 2022 (NNF22OC0078009), and the DTU Computing Center\cite{DTU_DCC_resource}.

\section{Acknowledgements}
We thank Prof.\ Peter Uhd Jepsen for providing the experimental THz pulse used in this work. 

\newpage
\bibliographystyle{elsarticle-num} 
\bibliography{bibi}
\appendix
\renewcommand{\thesection}{\Alph{section}}

\newpage
\section{The Fitting Procedure}\label{App:fitting}
An accurate description of the reservoir broadening functions $\mathbf{\Gamma}_\alpha(\epsilon)$ is important for describing the system's coupling to the electrode and it is furthermore necessary to accurately replicate steady state properties. A detailed fitting procedure that can be applied in the general case is given here. The goal here is to minimize the loss of structure in the $\mathbf{\Gamma}_\alpha$'s which will inevitably come from using the more restricted basis of Lorentzians.
\subsection{The cost function for optimizing the centres and widths of the Lorentzian basis functions.}\label{app:fit_1}
The cost function for the full problem is obtained by summing over the individual components and the starting point is therefore by fitting a scalar function on the real line to the form seen in Eq. \eqref{eq:GammaForm}. It is assumed that the function $f$ is sampled in points $\{x_i\}$.  An approximation of the function $f$ can then be given as the sum,
\begin{align}
    f(x) \approx \sum_{i}f(x_i)\phi_i(x),
\end{align}
with
\begin{align}
\phi_i(x) = \left\{
        \begin{array}{ll}
            \frac{x - x_1}{x_2 - x_1} & \quad x_2> x \geq x_1 \\
            \frac{x_2-x}{x_3 - x_2} & \quad x_3> x \geq x_2 \\
            0 &\quad \mathrm{else}
        \end{array}
    \right.
\end{align}
while the wanted expression is a function on the form\deleted[id=ABL3]{,} 
\begin{align}
    \mathcal{L}(x) = \sum_{l}^{N_L}\Gamma_{l}L_l(x).
\end{align}
Comparing the two expressions (real for now) with a $L^2$ norm, the norm of their difference can be written as
\begin{align}
    E &= \int_{-\infty}^{\infty}\!\!(f(x) - \mathcal{L}(x))^2\mathrm{d}x \\
      &= \int_{-\infty}^{\infty}\!\!(f(x))^2\mathrm{d}x + \int_{-\infty}^{\infty}\!\! (\mathcal{L} (x))^2\mathrm{d}x - 2\int_{-\infty}^{\infty}\!\!\! f(x)\mathcal{L}(x)\mathrm{d}x.
\end{align}
This is the type of expression which should be minimized in the end. In this simple case there are $3\cdot N_L$ unknowns to minimize the cost function $E$ with respect to. In the case of fitting a complex Hermitian matrix, there are roughly $n_o(n_o+1)\cdot  3\cdot N_L$ variables to minimize for. This can be drastically reduced by considering taking the Lorentzian poles $z_l = \epsilon_l +i\gamma_l$ as fixed. A linear system can then be obtained by setting $\frac{\partial E}{\partial \Gamma_j} = 0$:
\begin{align}
    \frac{\partial E}{\partial \Gamma_j} &= \frac{\partial\phantom{E}}{\partial \Gamma_j}\sum_{l l'}\Gamma_l \Gamma_{l'}\int_{-\infty}^{\infty}L_l(x)L_{l'}(x)\mathrm{d}x
    \\
    &- \frac{\partial\phantom{E}}{\partial \Gamma_j} \sum_{il}2f_i \Gamma_{l'}\int_{-\infty}^{\infty}\phi_i(x)L_{l'}(x)\mathrm{d}x\label{eq:dEdGamma}\\
    &=0
\end{align}
which implies
\begin{align}
     \mathbf{M}\vec{\Gamma} = \vec{v}\label{eq:GammaEquation}
\end{align}
with
\begin{align}
M_{ll'} =  \int_{-\infty}^{\infty}L_l(x)L_{l'}(x)\mathrm{d}x 
\end{align}
and
\begin{align}
    v_{l'} = \sum_i f_i \int_{-\infty}^{\infty}\phi_i(x)L_{l'}(x)\mathrm{d}x.
\end{align}

\begin{figure}[h]
    \centering
    \includegraphics[width=0.5\textwidth]{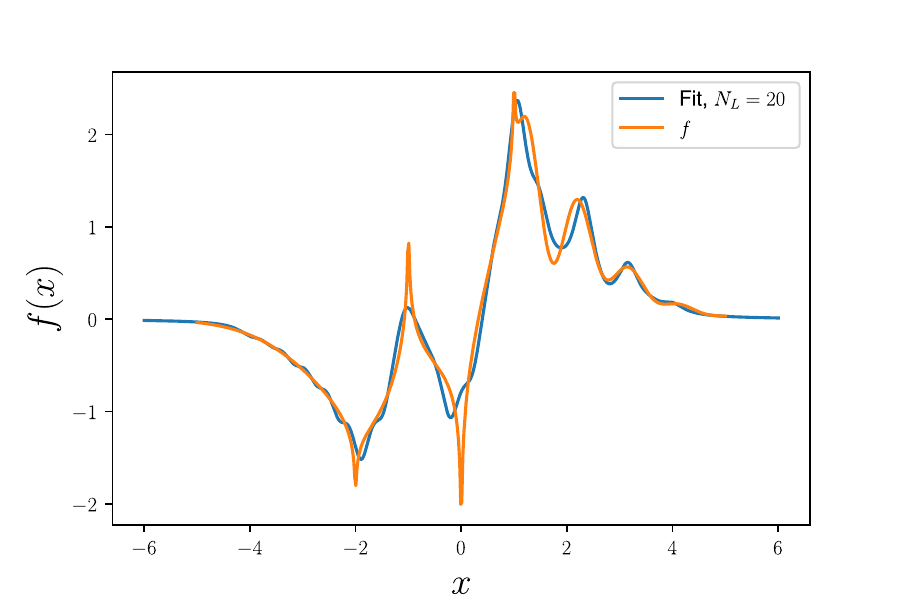}
    \caption{Demonstration of Eq. \eqref{eq:GammaEquation}. Function $f$ fitted using 20 Lorentzian basis functions on an equally spaced grid from  $x=-4$ to $x=4$ and $\gamma_i = \gamma = 0.25$. }
    \label{fig:SimpleFit}
\end{figure}
In Fig. \ref{fig:SimpleFit} a fit of a function $f$ is seen. The very sharp features of $f$ are hard to capture because of the basis not containing Lorentzians that are sufficiently narrow to fit the peak. Placing the poles and choosing their linewidths in a suitable way is part of the fitting procedure. 
\newline
This example does however show the coefficients, $\Gamma_l$, follows from the positions of the centres and widths of the Lorentzian functions. This reduces the number of variables needed to minimize w.r.t. to only $2N_L$. When minimizing the cost function $E$ one will get into trouble if one has added the same Lorentzian function to the basis twice, as $\mathbf{M}$ is singular because of the linear dependent basis. One can alleviate this by adding a cost term that drives the Lorenzians apart in the complex plane, a "repulsive term" so to speak. We therefore introduce an {\it ad hoc} term on the form,
\begin{align}
    R = \alpha_{PO}\sum_{l\neq l'}\tan(\frac{\pi}{2} O_{ll'}) \quad \mathrm{with}\quad O_{ll'} =  \frac{M_{ll'}}{\sqrt{M_{ll}M_{l'l'}}}.
\end{align}
Here $O_{ll'}$ is a normalised overlap that is always unity when two Lorentzians are on top of each other. The $\tan(\frac{\pi}{2}x)$ function is sharply divergent as $x\rightarrow 1$ giving the desired effect of punishing a linearly dependent basis. 

\subsection{Gradient of the cost-function}\label{app:fit_2}
What is left is to determine an expression for the gradient of the cost functions which will enable efficient minimization of the cost w.r.t. positions and broadenings, $\epsilon_i$ and $\gamma_i$, of the Lorentzian basis functions. Taking a derivative w.r.t. $x_i$ being either $\epsilon_i$ or $\gamma_i$, the following expression is obtained as,
\begin{align}
\label{eq:x_gradient_of_E}
    \frac{\partial E}{\partial x_i} =  \Gamma_{i}^2\frac{\partial}{\partial x_i}M_{ii} + \sum_{l\neq i}\Gamma_{i}\Gamma_{l}\frac{\partial}{\partial x_i}M_{il} + 2\sum_{j}f_j\Gamma_{i}\frac{\partial}{\partial x_i}K_{ji},
\end{align}
where $K_{ji}$ is defined as
\begin{align}
    K_{ji} = \int_{-\infty}^{\infty}\phi_j(x)L_i(x)\mathrm{d}x.
\end{align}
Both the $M_{ij}$ and $K_{ji}$ integrals can be evaluated analytically, making for an easily evaluated sum. The gradient of $R$ can also be evaluated using the chain rule and $\frac{\mathrm{d}\phantom{x}}{\mathrm{d}x}\tan(x) = 1/\cos^2(x)$ as,
\begin{align}
    \frac{\partial R}{\partial x_i} = \alpha_{PO}\sum_{l\neq l'}\frac{\pi/2}{\cos^2(\frac{\pi}{2}O_{ll'})}\frac{\partial O_{ll'}}{\partial x_i}.
\end{align}
The partial derivative$\frac{\partial O_{ll'}}{\partial x_i}$ can be evaluated numerically with finite differences. Summing up $E$ and $R$ and their gradients gives in the end a method for efficiently minimizing the error related to the Lorentzians. In practice, SciPy is used for this minimisation, as it has a large selection of algorithms to choose from\cite{2020SciPy-NMeth}. The extension to complex valued functions could in principle be done in a variety of ways, but here it is chosen to simply add the error on the real and the imaginary part of the fit.
\subsection{Limitations}
The two Subsections \ref{app:fit_1} and \ref{app:fit_2} are valid as long as one does not need to consider constraints on the function $f$. This is however the case when the matrix-valued level-width functions are considered. In particular the positive semi-definite condition is troublesome, as one has to verify the non-negativity of the eigenvalues of $\mathbf{\Gamma}_\alpha (\epsilon)$ for a dense grid of $\epsilon$. The fit $\mathbf{\Gamma}_\alpha$ of the level-width function will be approximate and therefore has regions of $\epsilon$ where $\mathbf{\Gamma}_\alpha$ has negative eigenvalues. One can however fit make the fit $\mathbf{\Gamma}_\alpha$ without initially constraining the fit, and then after the fitting process iteratively modify the weights until $\mathbf{\Gamma}_\alpha$ is positive semi-definite within a given tolerance. A scheme for iteratively making the fit $\mathbf{\Gamma}_\alpha$ positive semi-definite has been implemented in the "iterative\_PSD" method in the block\_sparse class of the Block\_matrices code.

\section{Identifying zero-elements of the various quantities}\label{sec:zeroele}

\begin{figure}
    \centering
    \includegraphics[width = 1\linewidth]{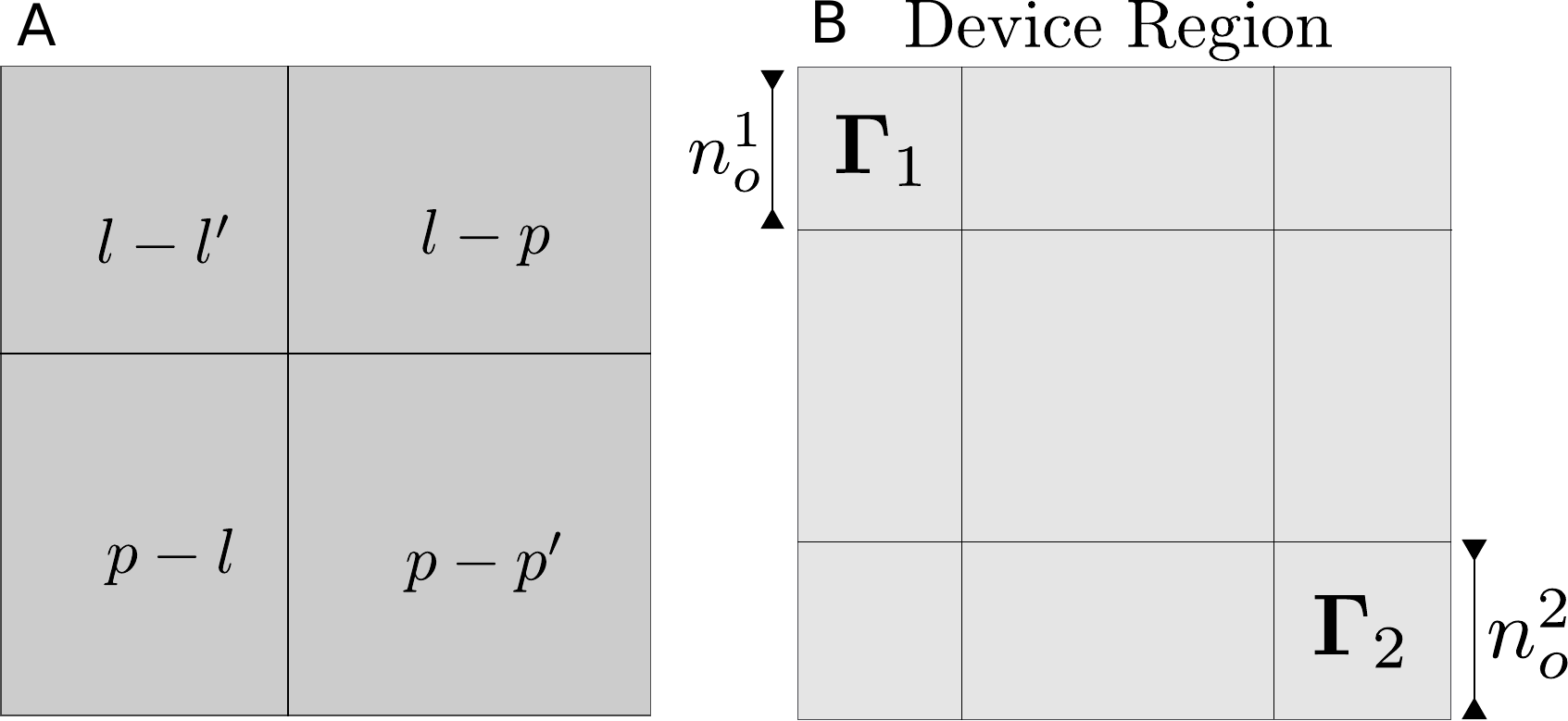}
    \caption{A) Arrangement of modes the two mode-types. $l$-index indicates pole from the Lorentzian fit, while $p$-indicates a pole from the Fermi-function. B) Matrix-structure of the $\mathbf{\Gamma}$'s in sorted LCAO basis. The $n_\alpha=2$ case.}
    \label{fig:Zeros}
\end{figure}
Firstly there are some guaranteed zeros, as have also been identified in Ref. \cite{Popescu2016}, $\Omega_{k,\alpha x c \alpha' x' c'}$ is zero when $x$ and $x'$  are both pole-indices of from the same Fermi-function. 
This is the $p-p'$ block in the Fig. \ref{fig:Zeros}A, which for a regular fit is going to be around 1/8  of the total elements of $\Omega_{k, \alpha x c\alpha' x' c'}$. Even through these are zero, they are still included in the array of $\Omega$ just because less function calls has to be made this way. \\
Next, the level-width function $\mathbf{\Gamma}_\alpha$ (in the non-orthogonal basis) is only nonzero inside the blocks of the device-region which couples to the $\alpha$'th electrode. This presence of zeros in most of the entries of the full $\mathbf{\Gamma}_\alpha$ matrices means that there is a number $n_o' \leq n_o$ that is an upper bound on how many non-zero eigenvalues the matrices $\mathbf{\Gamma}_\alpha(z_p)$ and $\mathbf{W}_{\alpha, l}$ have. The matrix structure of the $\mathbf{\Gamma}_\alpha$'s is seen in Fig. \ref{fig:Zeros}B, where one would choose $n_o' = \max(\{n_o^1, n_o^2\})$. This also generalises to more than two electrodes straightforwardly. If the eigenvalues of the $\Gamma_\alpha$'s are sorted according to absolute magnitude, then it is furthermore seen directly from Eq. \eqref{eq:PsiDef} that $\vec{\Psi}_{k,\alpha x c} = 0$ for $c>n_o'$. 
By inspecting Eq. \eqref{eq:omgEOM} it is seen that this also carries over to the $c$-indices of $\Omega$, meaning $\Omega_{k, \alpha x c\alpha' x' c'} (t)= 0$ if $c>n_o\vee c'>n_o'$.  Both of these facts can also be derived from just considering the ODE system  Eqs. \eqref{eq:psiEOM} and \eqref{eq:omgEOM} together with the initial conditions.
The effect of this is that for any system, there are a significant part of the quantities with orbital index $c > n_o'$ that are zero, and thereby significant amounts of computational and memory cost can be avoided. This property also carries over to the calculation of $\mathbf{\Pi}_\alpha$ in Eq. \eqref{eq:PIfromPSI}. $\Omega_{k,\alpha x c, \alpha'x'c'}$, which is the largest array from a memory point-of-view, therefore only scales with the couplings to the environment squared.\\
Lastly note that the $\mathbf{\Gamma}_\alpha$, on the form depicted in Fig. \ref{fig:Zeros}B may be represented in a non-orthogonal basis, necessitating a transformation into an orthogonal basis. We can consider a general unitary transformation and how it affects $\mathbf{\Gamma}_\alpha$
\begin{align}
    \mathbf{\Gamma}_\alpha ' = i (\mathbf{U}^\dagger\mathbf{\Sigma}_\alpha \mathbf{U} - \mathbf{U}\mathbf{\Sigma}_\alpha^\dagger \mathbf{U}^\dagger)
\end{align}
which is not necessarily the same as $\mathbf{\Gamma}_\alpha \rightarrow \mathbf{U}^\dagger\mathbf{\Gamma}_\alpha \mathbf{U}$ unless $\mathbf{U}$ is also a hermitian matrix. Applying e.g. a Lowdin transformation satisfies this and can therefore be used interchangeably. This means fitting $\mathbf{\Gamma}_\alpha$ in the non-orthogonal, where much fewer matrix elements needs to be fitted and then transforming to the orthogonal Löwdin-basis can be done. Furthermore the number of non-zero eigenvalues are still the same before and after the transformation. Only the eigenvectors are transformed.
\section{Convergence Checklist}
\label{app:convergence}
Here we provide a list of essential convergence parameters one should check in order to have a physically sound calculation. 
\begin{itemize}

\item The steady-state calculation should be converged. Here the reader is referred to Refs. \cite{brandbyge2002density, papior2017improvements} and the TranSIESTA tutorials for the details of carrying out a steady-state calculation.
\item The energy window around the equilibrium Fermi energy on which we fit $\mathbf{\Gamma_\alpha}$ should extend well beyond peak values of the time-dependent chemical potentials ($\mu_\alpha + \Delta_\alpha(t)$).
\item The Lorentzian expansion of the level-width function (see Eq. \eqref{eq:GammaForm}) should both reproduce the reference steady-state transmission function well over the energy window, while at the same time be positive semi-definite.
\item Number of poles in the Fermi-function should be sufficient to describe the electrode filling over the entire interval where $\mathbf{\Gamma_\alpha}$ is nonzero.
\item The contour chosen for the SCF tool needs to converge the density matrix. Inspect the value of $\max{(|\frac{\mathrm{d}\bsigma^0}{\mathrm{d}t}|)}$ printed in the output of the psinought code. The printed value is the derivative of $\bsigma^0$ when computed with the pole-expansion from Eq. \eqref{eq:FermiFunc}.
\item The error tolerance given to the adaptive Runge-Kutta solver needs to be checked for convergence. 
\end{itemize}

\section{TBtrans / sisl / Transiesta interface}
TBtrans is currently a widely used transport code and can do many of the calculations of $\mathbf{\Sigma}_\alpha$ given the Hamiltonian of the system\cite{papior2017improvements}. The python-package sisl\cite{zerothi_sisl} extends the ability of TBtrans while simultaneously adding advanced features such as real-space self-energy calculators. However, for the purposes of the fitting code, a goal of containing most of the calculation to one notebook has been aimed for. A supplementary python package $\mathrm{siesta\_python}$ has been made for the purposes of keeping the manipulation of fdf-files out of the picture, only keeping geometry definition and manipulation, together with executing some simple functions which executes tbtrans, sisl and TranSIESTA in the background. This package serves as a general siesta-calculator object, being able to script and create DFT-calculations from defined geometries, and can also be given other instances of the class as electrodes for instance. The general transport setup Fig. \ref{fig:TransportSetup} is handled internally in this code and works also as a standalone code for scripting workflows of transport calculations. 

\section{Generalisation to superconducting device region} \label{sec:SC}
The EOM approach can be extended to a superconducting device region with a pairing field $\Delta$, described by the BCS Hamiltonian \cite{tuovinen2016time, bogoljubov1958new, nambu1960quasi} 
\begin{align}
    H_D^{BCS} = \sum_{ijss'}h_{ijss'}c^\dagger_{is} c_{js'}+ \sum_{ij}^{ss'}\left[ \Delta_{ij}^{ss'}c^\dagger_{is}c^\dagger_{js'} +\Delta_{ij}^{ss'*}c_{is}c_{js'} ]\right],
\end{align}
while the electrode Hamiltonian remains in the normal state i.e. without any pairing field. The inclusion of the pairing field can be accounted for by employing the Bogoliubov-de Gennes transformation, which in effect means substituting at each original matrix entry the $4\times 4$ matrix\cite{tuovinen2016time}
\begin{align}
    \mathbf{H}^{BdG}_{ij} = 
    \begin{bmatrix}
    h_{ij}^{\uparrow \uparrow}/2 & h_{ij}^{\uparrow\downarrow}/2 & \Delta_{ij}^{\uparrow\uparrow} & \Delta_{ij}^{\uparrow\downarrow}\\
    h_{ij}^{\downarrow \uparrow}/2 & h_{ij}^{\downarrow\downarrow}/2 & \Delta_{ij}^{\downarrow\uparrow} & \Delta_{ij}^{\downarrow\downarrow}\\
    \Delta_{ij}^{\uparrow\uparrow*} & \Delta_{ij}^{\downarrow\uparrow*} &-h_{ij}^{\uparrow \uparrow}/2 & -h_{ij}^{\uparrow\downarrow}/2\\
    \Delta_{ij}^{\uparrow\downarrow*} & \Delta_{ij}^{\downarrow\downarrow*} &  -h_{ij}^{\downarrow \uparrow}/2 & -h_{ij}^{\downarrow\downarrow}/2\\
    \end{bmatrix}
    \label{eq:Ham_bdg}
\end{align}
and also replacing the electrode level-width functions with
\begin{align}
    \mathbf{\Gamma}^{BdG}_\alpha(\epsilon) = \frac{1}{2}\begin{bmatrix}
     \mathbf{\Gamma}_\alpha(\epsilon) &0 & 0 & 0\\
    0 &  \mathbf{\Gamma}_\alpha(\epsilon) & 0 & 0\\
    0 & 0 & \mathbf{\Gamma}_\alpha(-\epsilon) & 0\\
   0 & 0 &  0 &  \mathbf{\Gamma}_\alpha(-\epsilon)\\
    \end{bmatrix}.
    \label{eq:Gamma_bdg}
\end{align}
Here the $\epsilon \rightarrow -\epsilon$ substitution is made such that the reflection happens around the chemical potential of lead $\alpha$. The Green's functions' EOMs and commutation rules, when the problem is formulated in Nambu space, are the same as for the normal case, leading to the same EOM eq. \eqref{eq:sigEOM}, eq. \eqref{eq:psiEOM} and eq. \eqref{eq:omgEOM}. The N-SC-N tool can be used to carry out the transformation prescribed in eqs. \eqref{eq:Ham_bdg} and \eqref{eq:Gamma_bdg} with a user-defined pairing field. Furthermore, BCS theory has recently been implemented in the SIESTA code\cite{reho2024density}.
\section{Implementation}
\subsection{{\tt zand} time-propagation code}
How to translate equations \eqref{eq:sigEOM}, \eqref{eq:psiEOM} and \eqref{eq:omgEOM} to an efficient computer program is covered here. The standard method for solving a first order ODE is by using a Runge-Kutta adaptive solver\cite{fehlberg1969low}. This method propagates the system in steps of d$t$ which are adaptively made smaller or bigger depending on how good two methods of different orders compares. The method boils down to evaluating the RHS of eqs. \eqref{eq:sigEOM}, \eqref{eq:psiEOM} and \eqref{eq:omgEOM} at $N$ points between $t$ and $t +\mathrm{d}t$. The basics of propagating an ODE on the form 
\begin{equation}
    \dot{\mathbf{y}} = f(\mathbf{y})
\end{equation}
with the RK method is the prescription for evaluating the solution value $\mathbf{y}(t+\mathrm{d}t)=\mathbf{y}_{i+1}$ by sampling $f(t,\mathbf{y})$ at intermediate steps $t_m = t +A_m \mathrm{d}t$, denoted $k_m$ and updating as
\begin{align}
    \mathbf{k}_m &= f(t_m, \vec{B}_{m}\cdot [\mathbf{k}_0, ...,\mathbf{k}_N, \mathbf{y}_i])\\
    \mathbf{y}_{i+1} &= \vec{CH}\cdot [\mathbf{k}_0, ...,\mathbf{k}_N, \mathbf{y}_i] \\
    \epsilon &= | \vec{CT}\cdot[\mathbf{k}_0, ..., \mathbf{k}_N] |
\end{align}
Here $A_m$, $\vec{B}_{m}$, $\vec{CH}$ and $\vec{CT}$ can be obtained using standard numerical approaches, e.g Refs. \cite{fehlberg1969low, dormand1980family}.

The Runge-Kutta solver for the ODE in equations \eqref{eq:sigEOM}, \eqref{eq:psiEOM} and \eqref{eq:omgEOM}  is parallelized in python using the mpi4py package, NumPy and Numba\cite{dalcin2021mpi4py,dalcin2011parallel, dalcin2008mpi, dalcin2005mpi, harris2020array, van1995python, lam2015numba}. A description of the program can be found in Fig. \ref{fig:Zand_mpi_structure}A and the basic MPI-structure can be seen in Fig. \ref{fig:Zand_mpi_structure}.

\begin{figure*}[h]
    \centering
    \includegraphics[width=1.0\linewidth]{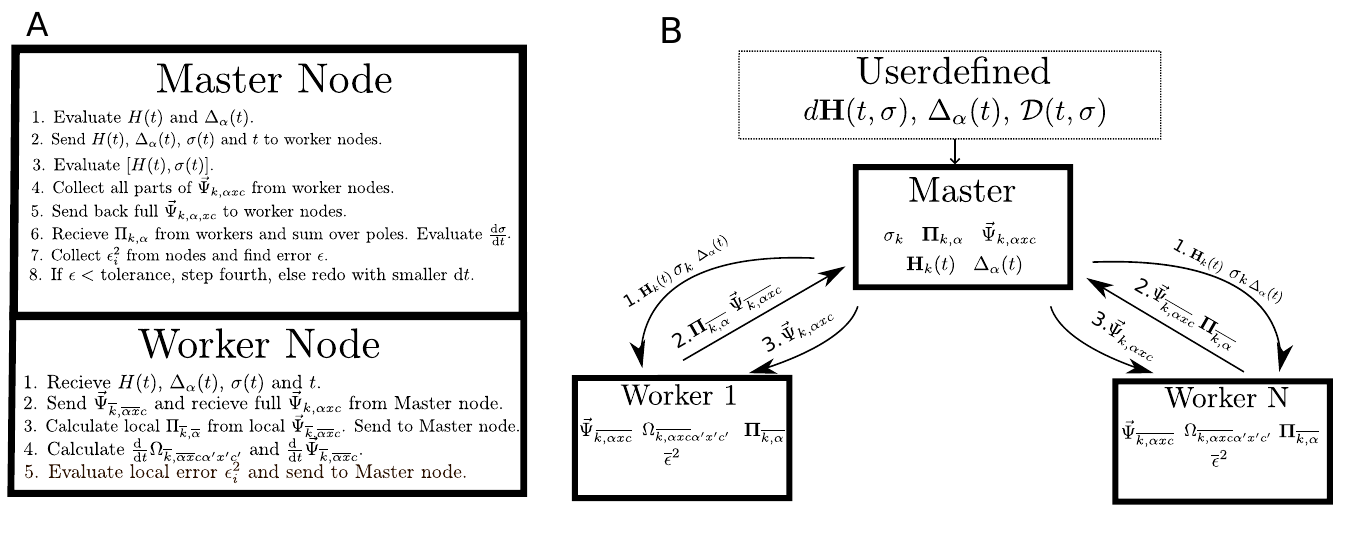}
    \caption{A) Recipe for efficient implementation of the auxiliary mode propagation scheme found in the \Packagename$\phantom{.}$code. Bared indices are the individual indices that has been distributed to the particular worker process. B) Diagram representing the MPI-node structure of the \Packagename$\phantom{.}$code. Bared indices are the individual indices that has been distributed to the particular worker process.}
    \label{fig:Zand_mpi_structure}
\end{figure*}
Each worker node is a separate mpi-process, but the operations taking place inside the workers can furthermore also utilize shared-memory parallelism. This is however up to the user to determine the best way of splitting the cpu-cores available between shared-memory threading and separate processes, and usually using a single thread per process is the best choice.

It can also be seen from Fig. \ref{fig:Zand_mpi_structure} that $\Omega_{k, \alpha x c \alpha' x' c' }$ does not need to be communicated between the processes. This is favorable because $\mathbf{\Psi}_{k, \alpha x c}$ is relatively inexpensive to transfer between mpi-processes compared to the whole $\Omega_{k, \alpha x c \alpha' x' c' }$ which has another set of ${\alpha x c}$ indices. The distribution of the bared indices to each worker can be done in a way that accounts for the computational burden associated with each. The master node continuously loops over the workers to check for data to collect.  
An important note on the implementation here is that $\mathbf{H}(t)= \mathbf{H}_0 + d\mathbf{H}(t,\sigma)$ can be made to depend on external code through e.g. the density matrix $\sigma$. This means the code can be linked with other DFT-codes that uses the LCAO approach. 
\subsection{{\tt psinought} code}
Equation \eqref{eq:Psi0eq} is equivalent to an inhomogeneous matrix equation on the form $\mathbf{A}\vec{\Psi} = \vec{b}$ for a "stacked" vector, $\vec{\Psi}$, written as,
\begin{align}
[\mathrm{Re}[\vec{\Psi}_{0,0,0}],..,\mathrm{Re}[\vec{\Psi}_{0,0,n_o'}], .., \mathrm{Im}[\vec{\Psi}_{0,0,0}],..,\mathrm{Im}[\vec{\Psi}_{n_\alpha, n_x, n_o'}]]
\end{align}
and $\vec{b}$ determined from the two first terms in eq. \eqref{eq:Psi0eq}. 
A modest system with two electrodes, a hundred modes, 25 couplings per electrode, and a hundred device orbitals will result in $\mathbf{A}$ being a $10^6\times10^6$ non-symmetric matrix. The sparsity of $\mathbf{A}$ depends of the number of couplings out of the device, but consists usually of more than 5\% nonzero elements, making both the construction unfeasible and a sparse solver inefficient. The system can instead be solved with a preconditioned iterative solver from SciPy\cite{sonneveld1989cgs, 2020SciPy-NMeth}. This method only requires a way to compute the matrix-vector product $\mathbf{A}\Vec{\Psi}$ and does not require explicit knowledge of the matrix elements of $\mathbf{A}$. The preconditioner is found to be crucial for obtaining the steady-state solution and can be obtained from the $\alpha x c$-diagonal part of eq. \eqref{eq:Psi0eq}. In brief the preconditioner is an approximation to the inverse and can be found by doing $n_k \cdot n_\alpha \cdot n_x\cdot n_o'$ matrix inversions of $(2\cdot n_o)\times(2\cdot n_o)$ matrices. Using it makes the iterative solver rapidly convergent.

Lastly, if a solution does not exist for eq. \eqref{eq:Psi0eq}, there is no steady state in the system under consideration, which is a warning sign of the underlying fit (coefficients in eq. \eqref{eq:GammaForm}) not being given enough care and that the $\mathbf{\Gamma}_\alpha$-matrices should be checked for negative eigenvalues on the real axis.

\subsection{SCF code}
The SCF code is implemented in along the same vein as TranSIESTA; using contour integration together with a integration over the bias window under nonequilibrium conditions\cite{papior2017improvements}. An adative integrator is used for this\cite{2020SciPy-NMeth}. It takes any density matrix based input that can be defined by the user and utilizes a Pulay type mixer from the sisl package\cite{zerothi_sisl} to determine the self-consistent solution.
\section{Expressions for $\mathbf{\Sigma}^0_\alpha$ and $\mathbf{\Sigma}^1_\alpha$ and their calculation.}\label{app:sig01}
Formulas for $\mathbf{\Sigma}^0_\alpha$ and $\mathbf{\Sigma}^1_\alpha$ are given as\cite{kwok2013time}
\begin{align}
	\mathbf{\Sigma}^1_\alpha &=\phantom{-} \mathbf{S}_{D\alpha}\mathbf{S}_{\alpha}^{-1}\mathbf{S}_{\alpha D}\label{eq:sig1}\\
	\mathbf{\Sigma}_\alpha^0(t) &= \mathbf{S}_{D\alpha}\mathbf{S}_{\alpha}^{-1} \mathbf{H}_{\alpha}^0 \mathbf{S}_{\alpha}^{-1}\mathbf{S}_{\alpha D} - \mathbf{S}_{D\alpha}\mathbf{S}_{\alpha}^{-1}\mathbf{H}_{\alpha D}^0-\mathbf{H}^0_{D\alpha}\mathbf{S}_{\alpha}^{-1}\mathbf{S}_{\alpha D}\\
    &= \Delta_\alpha(t)\left[ \mathbf{S}_{D\alpha}\mathbf{S}_{\alpha}^{-1} \mathbf{S}_{\alpha D} - \mathbf{S}_{D\alpha}\mathbf{S}_{\alpha}^{-1}\mathbf{S}_{\alpha D}-\mathbf{S}_{D\alpha}\mathbf{S}_{\alpha}^{-1}\mathbf{S}_{\alpha D}\right]\\
	&=-\Delta_\alpha(t) \mathbf{S}_{D\alpha}\mathbf{S}_{\alpha}^{-1} \mathbf{S}_{\alpha D} = -\Delta_\alpha(t) \mathbf{\Sigma}_\alpha^1
\end{align}

It is seen the inverse of $\mathbf{S}_\alpha$ is needed, which necessitates inverting the infinite matrix  $\mathbf{S}_\alpha$. This can however be done by recursive means as done with the surface Greens function \cite{sancho1985highly} and real space Greens function \cite{papior2019removing}, as it can simply be obtained by calculating a Greens function \replaced[id=ABL4]{$\mathbf{G}(z)$}{$G(z)$} with the (artificial) parameters $\mathbf{H}=0$ and $z=1$.

\section{Scaling}\label{app:scaling}
\added[id=ABL4]{The time-dependent part of the code scales with the number of orbitals in the device ($n_o$), the number of couplings to the electrodes ($n_c$), the number of poles ($n_x$) used in the expansion of the Fermi-function and level-width functions, the number of $k$-points and the number of timesteps needed. An estimate of the theoretical scaling when counting the (naive) number of operations needed for calculation of $\frac{\mathrm{d}\sigma}{\mathrm{d}t}$, $\frac{\mathrm{d}}{\mathrm{d}t}\vec{\Psi}_{\alpha x c}$ and $\frac{\mathrm{d}}{\mathrm{d}t}\Omega_{\alpha x c\alpha' x' c'}$ is seen in Fig. \ref{fig:fig_scaling_theoretical}. }
\begin{figure}
    \centering
    \includegraphics[width=1.05\linewidth]{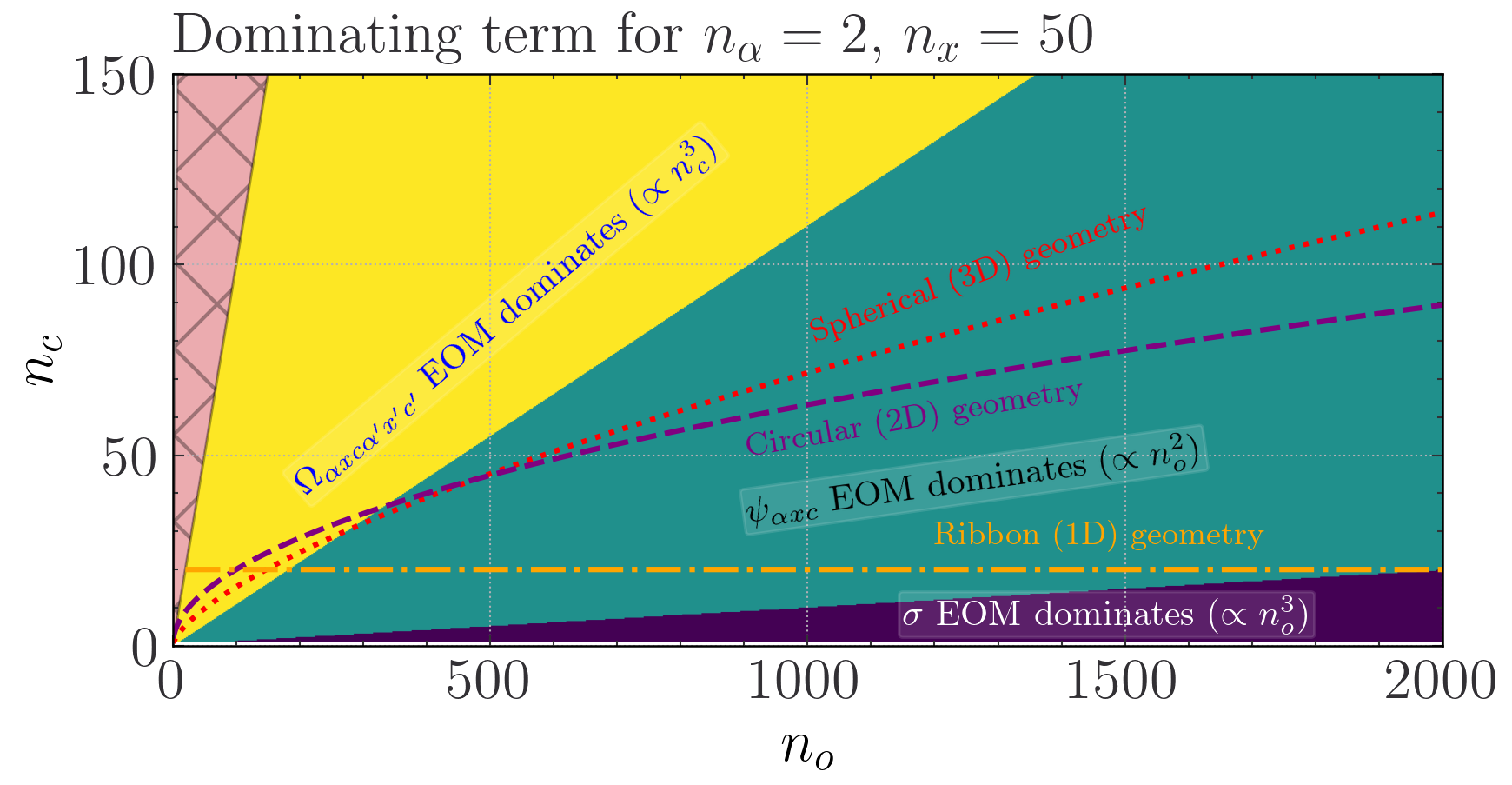}
    \caption{Regions colorcoded according to what term in the EOM that dominates. $n_\alpha$ and $n_x$ chosen for an example setup with 2 electrodes that has been expanded with 20 Fermi poles (that counts double) and 10 Lorentzians. Ideal scalings for spherical, circular and ribbon-like geometries are shown with dotted, dashed and dash-dotted lines respectively. Note that in the yellow region, computation time scales with $n_c$ not $n_o$.}
    \label{fig:fig_scaling_theoretical}
\end{figure}
\added[id=ABL4]{
It is seen that the $\Omega$ part of the EOMs dominate at small device sizes, and this part scales with $n_c$, not $n_o$. For intermediate device sizes, which are considered realistic for our purposes, the $\vec{\Psi}_{\alpha x c }$ EOM dominates and results in $n_o^2$ scaling in the run-time. It should be noted this $n_o^2$ scaling can be made linear in $n_o$ if the bandwidth of $\mathbf{H}(t)$ remains constant as $n_o$ increases. This is the case when using an orthogonal basis and a ribbon-like device. Since Zandpack is meant to work as an extension to TBtrans, the pivoting algorithms implemented in TBtrans can be carried over to the time-dependent propagation for evaluation of $\mathbf{H}(t)\vec{\Psi}_{\alpha x c}$ if an orthogonal basis is used. This feature is straight forward to implement and will be done in the future. Exploitation of a constant bandwidth in $\mathbf{H}(t)$ can also be done in the case of an orthogonal basis when the $\sigma$ EOM dominates. It should also be noted that Fig. \ref{fig:fig_scaling_theoretical} is sensitive to how the computation time is calculated. In practice matrix-matrix multiplication scales as $n_o^{2.807}$ instead of $n_{o}^3$ and this affects how soon the $\sigma$ EOM becomes the dominant part. Fig. \ref{fig:fig_scaling_theoretical} is however representative of the scaling using a nonorthogonal basis. Zandpack is however meant to be used in conjunction with DFT-codes where non-orthogonal bases are standard and therefore we will not give further treatment for this case. Lastly, the scaling with number of $k$-points and timesteps is trivial and linear. However, as the code uses an adaptive timestep, the number of timesteps will depend on the stepsize needed to propagate the system accurately. This will depend on the system under investigation. \\ }

\added[id=ABL4]{
A plot of run-time, normalized for number of time-steps, against the number of orbitals $n_o$ can be seen in Fig. \ref{fig:fig_scaling_for_reviewer}. It shows how the orbital-dependence affects the run-time for various device-widths $n_y$. A sub-linear scaling is seen in all cases when $n_o$ is small, which then turns into quadratic when the number of orbitals increases. Memory for computing the steady state eq. \eqref{eq:Psi0eq} becomes the limiting factor for how far out datapoints in Fig. \ref{fig:fig_scaling_for_reviewer} can be obtained. 
}
\begin{figure}
    \centering
    \includegraphics[width=1.05\linewidth]{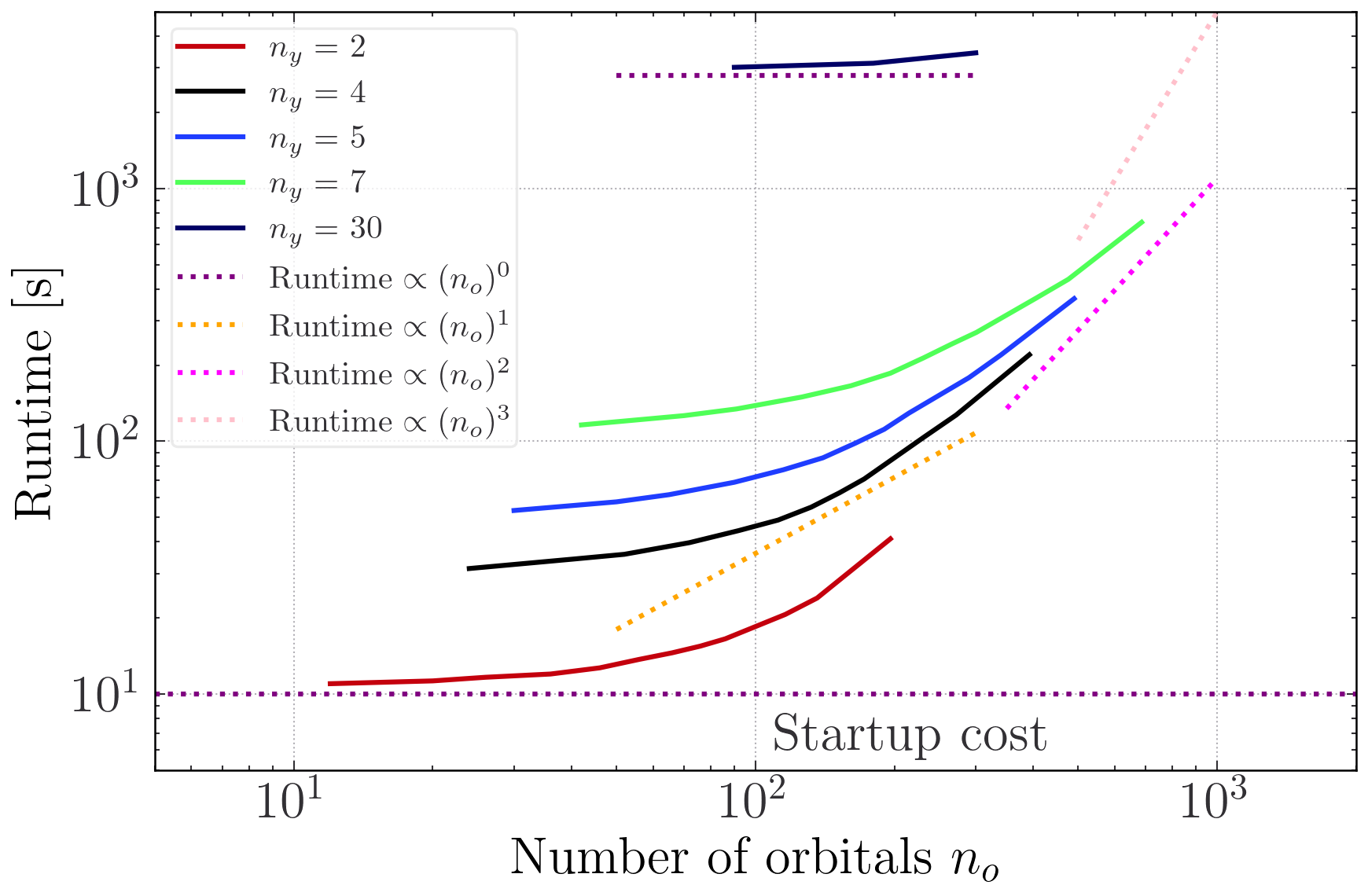}
    \caption{Measured runtime for various sizes of a rectangular ribbon geometry. $n_o = n_{reps, x} \times n_y$ and the number of couplings $n_c = n_y$ is equal to the number of repetions in the $y$-direction ($n_{reps,y} = n_y$). All results used between 850 and 900 steps to finish the propagation from $t_0=-25$fs to $t_1=100$fs.}
    \label{fig:fig_scaling_for_reviewer}
\end{figure}
\added[id=ABL4]{
The scaling with the number of processors are seen in Fig. \ref{fig:fig_scaling_for_reviewer_2}.  The runtime does benefit from more processes when going from 3 to 5 process, but it is only a improvement of around $19\%$, far from ideal. We suspect this is because the code is memory-bound in the case shown in Fig. \ref{fig:fig_scaling_for_reviewer_2}, else one should see better scaling. This is consistent with a very small additional improvement of $2\%$ when going from 5 to 7 processes, as is also seen in Fig. \ref{fig:fig_scaling_for_reviewer_2}.  }
\begin{figure}
    \centering
    \includegraphics[width=1.05\linewidth]{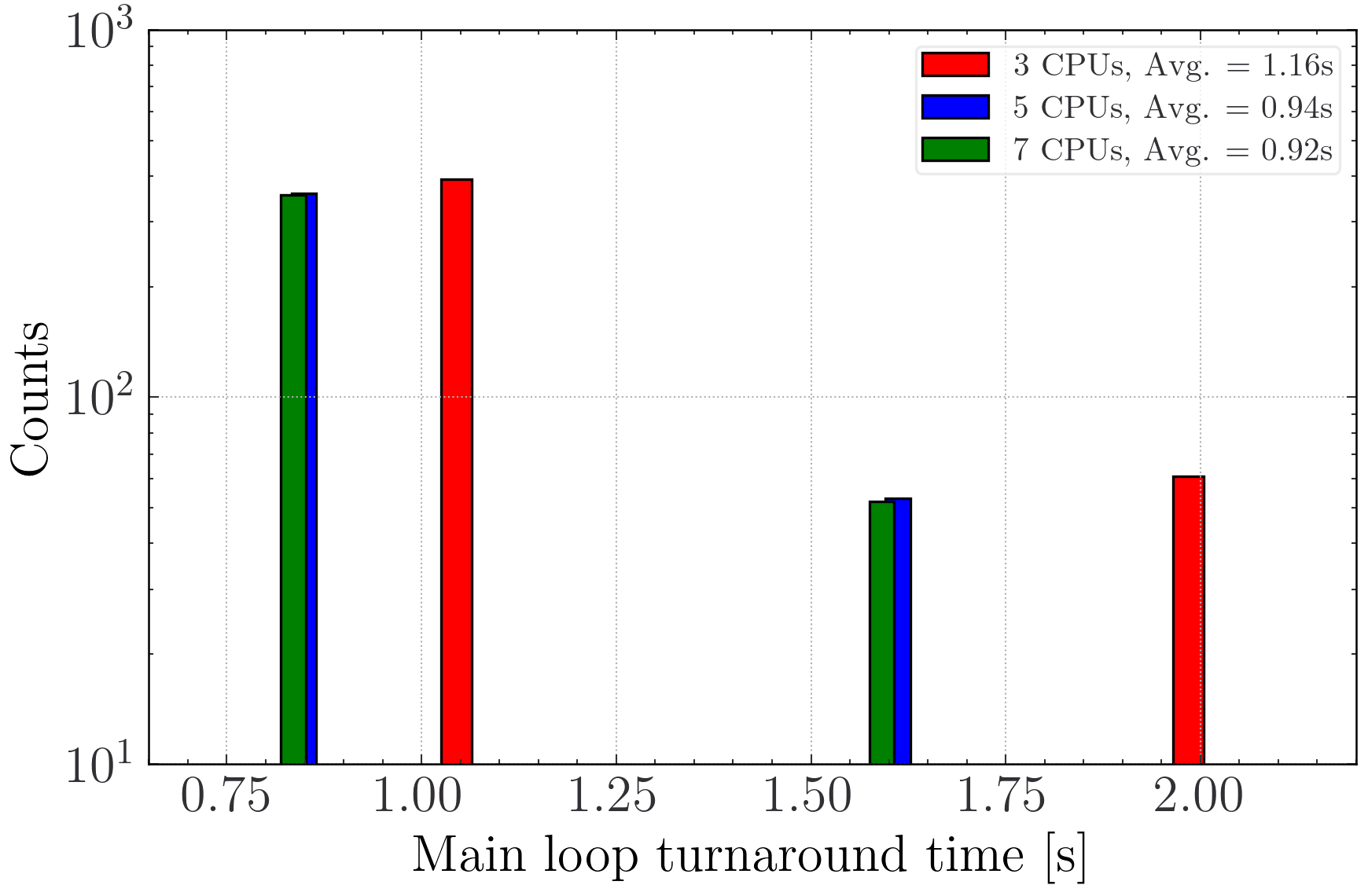}
    \caption{Main loop of code timed when running on 3, 5 and 7 CPUs, one process per CPU. Measured with $n_\alpha=2$, $n_x = 55$, $n_c=20$ and $n_o=460$. }
    \label{fig:fig_scaling_for_reviewer_2}
\end{figure}

\end{document}